\documentclass[fleqn,usenatbib]{mnras}

\usepackage{newtxtext,newtxmath}

\usepackage[T1]{fontenc}
\DeclareRobustCommand{\VAN}[3]{#2}
\let\VANthebibliography\thebibliography
\def\thebibliography{\DeclareRobustCommand{\VAN}[3]{##3}\VANthebibliography}

\usepackage{graphicx}	
\usepackage{amsmath}	
\usepackage{setspace}
\usepackage{todonotes}

\title[A New Monte-Carlo Radiative Transfer Simulation of CRSFs]{A New Monte-Carlo Radiative Transfer Simulation of Cyclotron Resonant Scattering Features.}

\author[Kumar et al.]{
Sandeep Kumar,$^{1,2}$
Suman Bala$^{2,3}$ \thanks{E-mail: sumanbala2210@iitb.ac.in } and
Dipankar Bhattacharya$^{2,4}$ \\
$^{1}$ Digambar Jain College, Dept. of Physics, Baraut, Baghpat, Uttar Pradesh 250611, India \\
$^{2}$Inter University Center for Astronomy and Astrophysics, Ganeshkhind, Post Bag 4, Pune 411007, India.\\
$^{3}$Indian Institute of Technology Bombay, Dept. of physics, Powai, Mumbai, Maharashtra 400076, India\\
$^{4}$ Ashoka University, Dept. Of Physics, Sonepat, Haryana-131029, India \\
}

\date{Accepted 2022 July 1. Received 2022 June 10; in original form 2022 March 14}

\pubyear{2021}

\begin{document}
\label{firstpage}
\pagerange{\pageref{firstpage}--\pageref{lastpage}}
\maketitle

\begin{abstract}
We present a new Monte-Carlo radiative transfer code, which we have used to model the cyclotron line features in the environment of a variable magnetic field and plasma density.
 The code accepts an input continuum and performs
only the line transfer by including the three cyclotron resonant processes (cyclotron absorption, cyclotron emission, cyclotron scattering). Subsequently, the effects of gravitational red-shift and light bending on the emergent spectra are computed.
 We have applied our code to predict the observable spectra from three different emission geometries; 1) an optically thin slab near the stellar surface, 2) an accretion mound formed by the accumulation of the accreted matter, 3) an accretion column representing the zone of a settling flow onto the star. Our results show that the locally emergent spectra from the emission volume are significantly anisotropic. 
However, in the presence of strong light bending the anisotropy reduces considerably. 
This averaging also drastically reduces the strength of harmonics higher than second in the observable cyclotron spectra. 
We find that uniform field slabs produce line features that are too narrow, and mounds with large magnetic distortions produce features that are too wide compared to the average widths of the spectral features observed from various sources. 
The column with a gently varying (dipole) field produces widths in the intermediate range, similar to those observed.
\end{abstract}




\section{Introduction }
Cyclotron Resonance Scattering Features (CRSFs) are quasi-harmonic
absorption and emission features found in the hard X-ray (10-100 KeV)
spectra of highly magnetized ($B\sim10^{12}G$) accreting neutron
stars. 
The CRSF was first observed in the X-ray spectra of Her-X1 \citep{1977MitAG..42..120T}.
 Till now, nearly 36 X-ray pulsars are known to exhibit cyclotron lines in their X-ray spectrum (see \citep{2019A&A...622A..61S} for a detailed review)
The observed cyclotron lines can be used to provide a direct estimate
of the field strength in the emission region, usually located close
to the neutron star surface, and help to probe the local magnetic field structure and accretion physics. 
The observed cyclotron lines are found in general to be broad, and
the fundamental usually has a shape more complex than a single Gaussian.
The higher harmonics have simpler profiles which can be usually well
approximated by a Gaussian.
The line energy of the fundamental CRSF ($\omega_{1}^{res}$)  is approximately given by the ``12-B-12 rule'' 
\begin{equation}
\omega_{1}^{res}\approx11.57\text{ KeV }B_{12}
\end{equation}

where $B_{12}$ represents the magnetic field in units of $10^{12}$
Gauss. 
The cyclotron line energies of the n'th harmonics ($\omega_{n}^{res}$) are slightly anharmonic due to relativistic corrections and are given by,
\begin{equation}
\omega_{n}^{res}(\mu_{i}^{'})=\dfrac{m(\sqrt{(1+2n(B/B_{c})\sin^{2}\theta_{i}^{'})}-1))}{\sin^{2}\theta_{i}^{'}}\label{eq:1.4}
\end{equation}
(In this work, all quantities are given in natural units, i.e. $\hbar=c=1$.) Here, $\theta_{i}^{'}$ is the angle between the photon
propagation direction and the magnetic field (\textit{B}), and $\mu_{i}^{'}=\cos\theta_{i}^{'}$. The mass of the electron is denoted by $m$, and $B_c$ ($=m^2/e=44.1\times10^{12}G$) is the critical magnetic field.
The resonance energy $\omega_{n}^{res}(\mu_{i}^{'})$ depends
both on the field strength $B$ and $\mu_{i}^{'}$.
The angle dependence of the resonance energy as well as the optical depth introduces a strong anisotropy in photon propagation and the emitted spectra. 

There is an extensive body of existing work on the formation of cyclotron
lines in X-ray binaries. Two approaches have normally been followed
to model the cyclotron lines: 1) solving the integro differential
equations of radiative transfer, 2) Monte Carlo simulations. The approach
of modeling of cyclotron lines by solving the radiative transfer equations
has been adopted in works such as \citet{Bonazola_et_al_1979, Meszaros_et_al_1980, Nagel_1980, Nagel_1981_a, Nagel_1981_b, Mesaros_Nagel_1985, Alexander_et_al_1989, Alexander_meszaros_1991, Bulik_et_al_1992, Bulik_et_al_1995} and \citet{Nishimura_2003, Nishimura_2005, Nishimura_2008, Nishimura_2011}. 
Monte-Carlo simulation is the other approach used
in the modelling of the cyclotron lines. The existing work in this category
include those by \citet{Yahel_1979, Pravdo_Bussard_1981, Araya_Harding_1996, Isenberg_1998_a, Isenberg_1998_b, Freeman_et_al_1999, Araya_et_al_1999, Schonherr_et_al_2007} and \citet{2017A&A...597A...3S, 2017A&A...601A..99S}. Among these the most relevant
for our work are the ones by \citet{Araya_et_al_1999, Schonherr_et_al_2007, 2017A&A...597A...3S, 2017A&A...601A..99S}, and \citet{Nishimura_2003, Nishimura_2005, Nishimura_2008,Nishimura_2011}.

\citet{Araya_et_al_1999} [hereafter AH99], for the first time, included
fully relativistic cross-sections and transition rates in their Monte-Carlo
simulations. They have produced a new set of results for near critical
magnetic fields using the relativistic cross-sections derived by \citet{Sina_1996}. Their simulations were performed for uniform field and uniform
density. The effect of optical depth and geometry on the cyclotron
line features were studied in this work. Their model produced narrow
line features. 

\citet{Schonherr_et_al_2007} [hereafter Schon07] used the code of AH99, and
developed a fittable model for the well known X-ray spectral analysis package, XSPEC.
They studied the effects of different geometry, optical depth, plasma
temperature, linear gradient in the magnetic field and photon angular
re-distribution. They also developed a Green's function technique
using which the spectra for different injected continuum shapes could
be obtained from the results of a single simulation with a flat continuum
input.

\citet{Nishimura_2003, Nishimura_2005, Nishimura_2008, Nishimura_2011} has investigated a number of issues
regarding the cyclotron line formation. \citet{Nishimura_2003} studied the
effect of dipolar magnetic field on the cyclotron features. \citet{Nishimura_2005}
studied the effect of a vertical field gradient, including both dipolar
and a crust-anchored multipolar \citep{Gil_et_al_2002} component. \citet{Nishimura_2008} studied the mechanism of formation of broad and shallow lines
as observed in most cyclotron line sources.
He used the geometry
of a stack of multiple sections, each with its own value of magnetic
field, density and temperature.
Finally composite spectra were presented
for both slab and cylindrical geometry, showing the complexities resulting
from multi-zoning.

The most recent work in Cyclotron Line MC simulation has been done by \citet{2017A&A...597A...3S, 2017A&A...601A..99S} [hereafter SC17] in a series of two papers.
They have
computed and stored interpolation tables needed for their MC simulation
code using the fully relativistic  photon-electron partial
scattering cross section from \citet{Sina_1996}.  These
interpolation tables have been used by their MC code to save computation time. They have done the cyclotron
line simulation for a cylindrical geometry, including parameter
gradients for density, magnetic field, parallel electron temperature
and matter velocity.
They have also incorporated different types of photon sources, namely  i) point
source ii) line source iii) plane source relative to magnetic field
direction at the base of column. In their simulation the
continuum of any shape can be generated
using the Green's function approach. 
They have also fitted the NuSTAR spectra of Cep X-4 with their new
XSPEC model \texttt{cyclofs} assuming a slab geometry and a FD-cut continuum. They found that the estimated
magnetic field differs significantly from the estimated value form a simple Gaussian absorption line. 

 \citet{Mukherjee_and_Bhattacharya_2012} [hereafter MB12] have estimated the field distortion in the polar cap accretion
mound, and studied its effect on the fundamental cyclotron feature.
They first solved the Grad-Sharanov equation for an axisymmetric magneto-static
equilibrium to estimate the magnetic field distortion for different
mound heights corresponding to different mass loading on the field
lines. The distorted field structure in the mound were then used to
phenomenologically compute the expected shape of the fundamental feature.
Light bending effects were included but no radiative transfer was
performed. The spin phase dependence of the line feature was investigated.
It was found that with the increase of mound height and hence of field
distortion, the fundamental feature tends to broaden and develop a
double-peaked structure. Strong angle averaging due to light bending
tends to wash out the angle dependence of the spectrum seen by a distant
observer. Little or no spin phase dependence is left in the spectrum
unless there happens to be dissimilar contributions from the two opposite
poles.

Any model that attempts to explain the observed features of cyclotron
lines such as large widths, variation of line energy and line width
with luminosity, anharmonic line ratios etc, must incorporate realistic fields, plasma density and emission geometries.
In earlier theoretical models one of two extreme emission geometries are considered: a fan beam or a pencil beam, but there
are sources with intermediate geometries where significant
contributions may come from both fan beam and pencil beams \citep{Becker_et_al_2012}. No attempt till date has been made to incorporate
both the geometries in a model. We have developed a Monte-Carlo code
CLSIM in which arbitrary magnetic field and plasma density variation
can be incorporated. The fan beam and pencil beam geometries can also
be simultaneously incorporated. Many of the basic techniques in our
code are based on the Monte-Carlo implementation of AH99 but we differ
from AH99 in those parts of the implementation where non-uniform magnetic
field and plasma density are incorporated. Additionally, light
bending effects are also incorporated in our scheme. Finally we have
implemented a scheme of weighting entire photon trajectory trees as
a function of the initial injected photon energy, allowing us to compute
the resulting spectra from an input continuum of any shape, using a
single initial simulation with a flat continuum. 

\section{Physical assumptions of our model\label{sec:sec2}}
In this work, we solve the radiative line transfer problem with inputs such as geometry, magnetic field structure, plasma density, temperature and velocity profiles from previous works (\citet{Becker_and_wolf_2007}, MB12, AH99).
In very strong magnetic fields ($\approx10^{12}G$) the problem of line transfer is challenging \citep{Wang_et_al_1988}, so, we have made some simplifications and approximations.
The physical assumptions and approximations made in our
work are as follows;

\begin{itemize}
\item \textbf{Static Column}: We assume that the plasma inside the accretion column is in a magnetostatic configuration. This may apply to a slowly sinking plasma in the accretion column. 

\item \textbf{Resonance processes}: We are mainly interested in the modelling of CRSF features so only the line transfer is considered. Modification of the injected continuum is considered solely due to the three resonant processes (cyclotron absorption, scattering and cyclotron emission). 
 The production and absorption of continuum photons during the radiation diffusion inside the simulation volume have been ignored. 

\item \textbf{Plasma thermodynamic state}:
We consider a low density plasma ($n_{e}\sim10^{22}$ cm$^{-3}$), and electrons are assumed to be in the ground state of Landau levels $n=0$ since for high magnetic fields ($\sim10^{12}$ G) the cyclotron decay rates dominate over collisional excitation rates (AH99, \citet{Schonherr_et_al_2007}). 
We assume that the plasma is in statistical equilibrium and is defined by a parallel electron temperature $T_{e}$ and a relativistic energy distribution; \\
\begin{equation}
f_{e}(p)dp=N\exp\left(-m\dfrac{\sqrt{1+p^{2}}-1}{T_{e}}\right)\, dp\label{eq:2.16}
\end{equation}

Here $p$ is the momentum parallel to the magnetic field and $T_{e}$
is the temperature characterizing the motion of the electrons parallel
to the magnetic field. 
Although the temperature may vary in the accretion column (\citet{Isenberg_1998_a, Isenberg_1998_b, Nishimura_2011}, AH99) in this work we have assumed it to be constant. We have conducted our simulations for different values of $T_{e}$.
As the electrons are assumed to be in the Landau ground state with zero perpendicular momentum, the distribution given in Eq--\ref{eq:2.16} is one dimensional. 

\item \textbf{Optical depth}: Previous Monte-Carlo simulations for modelling the cyclotron line are restricted to the optical depth in the range ($10^{-4}-10^{-3})$, as the computation time is always an issue with Monte-Carlo simulations when it deals with every single microscopic event. In this work, we have used the relativistic magneto-Compton cross-section by \citet{Sina_1996} which is also very computation intensive. So, considering available computational resources, we have restricted our simulations to regions of maximum Thompson optical depth of $\sim10^{-3}$, which corresponds to a line centre optical depth of $>10^{2}$ at the fundamental. Although higher optical depth can produce a more accurate shape of the CRSFs, it is not achievable as computation time increases quadratically with $\tau$ (as the number of scatterings increases as $\tau^{2}+\tau$).

\item \textbf{Magnetic field}: In our simulations, we have explored the magnetic field in the range $(0.84\times10^{12}-6.6\times10^{12})\text{G}$, which covers most of the observed range of cyclotron lines. 

\item \textbf{Photon polarization modes and electron spin state}: The magnetized vacuum photon polarization modes, appropriate for the high field, low-density plasma are used in computing the scattering cross-section. In this regime, it is adequate to consider polarization averaged radiative diffusion, which we adopt in the computations present in this work.
We adopt the best possible choice of spin states and polarization states from \citet{Sokolov_Ternov_1968} and \citet{Shabad_1975} respectively.

\item \textbf{Cross-sections and decay rates}: 
The relativistic cross-sections \citep{Sina_1996, Harding_Daugherty_1991} and transition rates \citep{Harding_Preece_1987} are considered in this work.

\end{itemize}

\section{A new Monte Carlo code CLSIM\label{sec:sec3}}

Based on the assumptions stated above, we have developed a complete Monte-Carlo radiative transfer code to model resonant cyclotron processes including absorption, emission and magneto-Compton scattering. 
While we use some of the techniques outlined in AH99, the implementation is entirely our own. One of the recent works on the Monte-Carlo simulation of cyclotron line is SC17.
A comparison of this work with AH99 and SC17 is given bellow. \par
We have computed the line profiles including up to ten harmonics, while AH99 included up to four, and SC17 have included
up to five harmonics. In
SC17 and this work the angle of scattering has been selected using the differential scattering cross-section by \citet{Sina_1996} while AH99 have used the differential transition rates by \citet{Latal_1986}. 
We have also included additional features that enhance the applicability of our code. 
We perform our computation in real space instead of optical depth space as done by AH99 \& SC17.
 SC17 and our work include the effects of gravitational red-shift, but only our code includes the effect of light bending, allowing us to generate results that can be directly compared with observations. 
In this work, we implement a scheme of weighting entire photon trajectory trees as a function of the initial injected photon energy, allowing us to
compute the resulting spectra from an input continuum of any shape,
from a single initial simulation with a flat continuum, whereas SC17 has used the Green's function approach for the same.
In SC17 the simulation volume has been split into different cylinders with arbitrary dimensions, whereas we have split the simulation volume
into small cells.
AH99 and SC17 have considered a constant magnetic field and plasma density within the simulation volume, but our code is capable of including any arbitrary spatial variation of the magnetic field and plasma parameters within the simulation volume.
In this work we have solved the radiative diffusion problem for three applications
cases, as mentioned in Table \ref{tab:4.1}. \par
\subsection{Application cases}

We use our Monte Carlo code to compute the emergent spectra in three
broadly defined cases (Table--\ref{tab:4.1}) mentioned below;

\begin{table}
\renewcommand{\arraystretch}{1.1}
\centering
\caption{Application cases of our code.}
\begin{tabular}{|p{1.cm}p{1.5cm}p{1.cm}p{1.5cm}p{1.cm}|}
\hline 
Type & Geometry & Magnetic field structure  & Density variation  & $T_{e}$ \\
\hline 
\hline 
column \par

Case-I  &  Slab10   &  Uniform  &  Uniform  &  Uniform \\
\hline 
 mound \par

 Case-II &  Slab10  &  MB12  &  \citet{Paczinsky_1983} &  Uniform \\
\hline 
Accretion Column & Cylinder with  & Dipolar & \citet{Becker_and_wolf_2007} & Uniform \\
Case-III & supercritical luminosity & & \\
\hline \hline
\end{tabular}

\label{tab:4.1}
\end{table}

\begin{itemize}
\item CASE-I: An isothermal static plane circular slab 1-0 (illuminated from below) having a uniform magnetic field and uniform density. In the case of low luminosity sources the inflow of plasma is stopped
near the stellar surface generally via Coulomb collisions. 
In these situations slabs with scale height of $z_{c}\sim200\text{ cm}$ are expected \citep{Meszaros_et_al_1983, Harding_et_al_1984}. So, we have taken a similar emission geometry  with a plane circular
slab 1-0 (illuminated from below) of height $z_{c}=100\text{ cm}$
and radius $r_{c}=1\text{ Km}$. The simulation is performed for uniform
field, uniform plasma density and uniform temperature in the simulation
region. 

\item CASE-II:  An optically thin top layer of the accretion mound approximated as an isothermal static plane circular slab 1-0, with a distorted magnetic field, taken from MB12 and a density profile obtained using the equation of state given by \citep{Paczinsky_1983}. The accreted matter at the base of the accretion column is assumed to form a magnetically confined plasma mound.
The heavy mass loading at the base of the mound can severely distort the dipolar magnetic field structure (\citet{2004MNRAS.Payne&Melatos.351..569P}; MB12). Cyclotron lines which originate
in optically thin layer of accretion mound bears the signature of
distorted magnetic fields and can be used to
measure the extent of magnetic field distortion. \\
We follow the approach of MB12 to solve for the structure of the static
axisymmetric polar mound from the 2D Grad Shafranov equation using our choice of equation of state. The
solution is obtained in terms of the magnetic flux function $\psi(r,z)$
in cylindrical coordinates, from which the magnetic field may
be computed as :\\
\[
B_{r}=-\dfrac{1}{r}\dfrac{\partial\psi(r,z)}{\partial z},\,\,\,\,\,\, B_{z}=\dfrac{1}{r}\dfrac{\partial\psi(r,z)}{\partial r}
\]
We construct mound structures for different heights (e.g. 45m, 55m, etc).\\
In this case we simulate the cyclotron line formation in the optically
thin top layer of the plasma mound, which is approximated as a 1-0 plane circular slab geometry, with the distribution of magnetic
field and density as obtained from the GS solution. 
The Cartesian components of the magnetic field are obtained for
any point ($r$,$\phi$,$z$) in the mound as \\
\[
\begin{array}{l}
B_{x}=B_{r}\cos(\phi)\\
B_{y}=B_{r}\text{\ensuremath{\sin}(\ensuremath{\phi})}\\
B_{z}=B_{z}
\end{array}
\]
We model this thin layer as a slab of uniform thickness, with continuum
Thompson optical depth $\tau_{T}=10^{-3}$. The density as a function
of depth $l$ in this slab works out to be\\
\begin{equation}
\begin{array}{c}
n_{e}(l)=(X_{F}(l)/A_{F})^{3}/m_{He}\\
\\
X_{F}(l)=\dfrac{15}{16\sqrt{2}}\sqrt{\left(\xi^{2}(l)-\dfrac{8}{3}+\xi(l)\sqrt{\left(\dfrac{16}{9}+\xi^{2}(l)\right)}\right)}\\
\\
\xi(l)=80.89959983(l/L_{0})+1
\end{array}\label{eq:4.1.5}
\end{equation}
from the Grad Shafranov solution. where $L_{0}=1\text{Km}$ is a scaling
parameter, and plasma is assumed to be a electron-Helium mixture. 

\item Case-III: In this case, we deal with the situation at the opposite
extreme, where the accretion rate is high and the accretion column
is optically thick. In high luminosity sources, the flow of material
near the neutron star is decelerated by radiative shocks a few kilometers
away from the surface and an elongated column is formed in which the matter subsonically settles and generates radiation. For this case, we
place the source plane at a fixed optical depth inside the boundaries
of the cylindrical column. The density (Eq--\ref{eq:4.1.6}) and hence the
optical depth across the column increases towards lower altitudes.
The width of the optically thin layer ($\tau_{T}=10^{-3}$) is found to be very
small (millimeters) near the base of the column but much larger at
higher altitudes due to decreasing density. The injection surface
then can be approximated as a truncated cone buried inside the cylindrical
column (Fig--\ref{fig:f1}). The injected photons diffuse in the region bounded
by the injection surface and the outer surface of the cylindrical
column. This type of geometry can be considered as a cylindrical shell,
in which the radiation escapes sideways and also from a slab which is situated at
the top. The height and the radius of the accretion column are assumed to be
$h_{c}=\text{ 1 km}$ and $r_{c}=1\text{ km}$ respectively.
The magnetic field in the column is assumed to be dipolar, with a
strength $B^{'}=0.03$ (all the magnetic field values in this work are given in the unit of $B_c$). at the base. For distribution of plasma density
in the column we adopt the profile derived by \cite{Becker_and_wolf_2007}
for the case of super-Eddington luminosity, \\
\begin{eqnarray}
n_{e}(z) & = & \left(\dfrac{\dot{M_{c}}}{\pi r_{c}^{2}|v_{ff}(1-(7/3))^{-z/z_{c}}|}\right)/m_{He}\label{eq:4.1.6}
\end{eqnarray}
\begin{equation}
v_{ff}=\sqrt{\dfrac{2GM_{*}}{R_{*}}}\label{eq:4.1.7}
\end{equation}
where $\dot{M_{c}}$,$v_{ff}$ ,$z_{c}$,$r_{c}$ are the accretion
rate, free fall velocity, height of the accretion column and radius
of the accretion column respectively. $M_{*}$, $R_{*}$ are the mass
and the radius of the neutron star. \\

\end{itemize}

In all cases, we explore the line forming region with a maximum Thomson optical depth of $\tau_{T}\sim10^{-3}$.

\subsection{Our radiative transfer scheme\label{sub:radtran1}}
In our Monte-Carlo radiative transfer scheme we perform the polarization
averaged radiative transfer in the following steps.
\begin{enumerate}
\item A photon is injected at a location ($x_{inj}$,$y_{inj},z_{inj}$)
at the source plane with energy $\omega_{inj}$ in the direction ($\theta_{inj}$,$\phi_{inj})$
(Sec.\ref{sub:chap4photinject}). 
\item After injection, the propagation length of the photon before any scattering
or absorption is determined by a chosen free path $\lambda(B^{'},\omega_{i},\mu_{i})$. 
The photon either escapes, or is absorbed or scattered after propagating
the distance $\lambda$.
The probability of scattering and absorption
are computed and one from absorption or scattering is selected. 
\item For scattering or absorption the parallel momentum $p_{i}$ of the
electron is selected. 
\item The electron is excited to a higher state $n_{f}$ either through
scattering or through absorption. If absorption occurs then the photon
trajectory is terminated, if scattering occurs then the angle of scattering
$\theta_{f}$ is selected. 
\item The electron de-excites via radiative transitions to a lower Landau
level $n_{f}^{'}$ and generates a transition photon of energy $\omega_{t}$
and emitted in  direction ($\theta_{t}$,$\phi_{t}$). The electron
continues to de-excite and emit transition photons until it reaches
the ground state. All these transition photons are again injected
into the Monte-Carlo scheme. This process of photon propagation continues
until photons escape from the boundary of the simulation region. 
\item Finally the light bending effects are incorporated.
\end{enumerate}
It is to be noted that our Monte-Carlo scheme is based on AH99 but
with some differences which are necessary for the implementation of
varying magnetic field and varying density. In the rest of this section
we only mention those issues which are either important to highlight
or differs form AH99 scheme and related to our implementation only.
For full implementation of Monte-Carlo scheme using relativistic cross-sections
for uniform magnetic field and uniform density, we strongly refer the
reader to AH99.

\subsection{Photon Injection\label{sub:chap4photinject}}
The first step in the Monte-Carlo simulation is to inject the photons
from a prescribed continuum with an energy $\omega_{inj}$, propagating
in some direction ($\theta_{inj},\phi_{inj}$) from some position
($x_{inj},y_{inj},z_{inj}$) on the source plane. The selection procedure
of the energy of a photon at injection is explained in the next section
and position of photon injection and the direction of its propagation
are explained in Sec.\ref{sub:4.2.2}

\subsubsection{Selection of energy for photon injection\label{sub:4.2.1}}

The input continuum spectrum may be simulated by either drawing the
photon energies from a prescribed spectral distribution, or by first
carrying out the Monte Carlo simulation for an uniform photon energy
distribution, and later assigning weights to each emergent photons
as a function of its input energy and the prescribed continuum shape.
The latter method avoids the need to run the simulation separately
for each continuum shape, and is therefore our method of choice. For
each escaping photon, the escape location $x_{e},y_{e},z_{e}$, the
energy $\omega_{e}$ and the propagation angle $\theta_{e}$,$\phi_{e}$
and its mother photon parameters ($x_{inj},y_{inj},z_{inj}$,$\theta_{inj},\phi_{inj}$,$\omega_{inj}$)
are stored. The emergent spectrum arising out of a specific injected
continuum shape $f_{p}(\omega)$ may then be obtained by multiplying
a weight proportional to $f_{p}(\zeta w_{inj})$ to each escaping
photon, where $\zeta $ is the redshift of the mother photon with
energy $w_{inj}$. The output spectrum is then compiled by counting
the total weight of the escaping photons in different energy and angle
bins: 

\begin{equation}
N_{kl}(\omega_{k},\mu_{l})=A{\displaystyle \sum_{\begin{array}{c}
{\scriptscriptstyle |\omega_{e}-\omega_{k}|\leq\Delta\omega_{e}}\\
{\scriptscriptstyle |\mu_{e}-\mu_{l}|\leq\Delta\mu_{e}}
\end{array}}f_{p}(\zeta \omega_{inj})}
\end{equation}
where A is a normalization factor, which we have set to unity as we
are interested only in the relative shape rather than the total energy
in the spectrum. Apart from the flat continuum, we use the HCUT (High energy cutoff
power law) continuum shape in this work.

\begin{equation}
f_{p}(\omega)\varpropto\left\{ \begin{array}{c}
\omega^{-\Gamma-1}\,\,\,\,\,\,\,\,\,\,\,\,\,\,\,\,\,\,\,\,\,\,\,\,\,\,\,\,\,\,\,\,\,\,\,\,\,\,\,\,\,\,\,\,\,\,\,\,\,\,\,\text{ for }\mbox{\ensuremath{\omega}}\leq\omega_{cut}\\
\omega^{-\Gamma-1}\exp\left(-\dfrac{\omega-\omega_{cut}}{\omega_{fold}}\right)\text{ for }\omega>\omega_{cut}
\end{array}\right)\label{eq:6.2.3.1}
\end{equation}

\subsubsection{Selection of angle and position co-ordinates for photon injection\label{sub:4.2.2}}

The method of selection of position and angle of photon injection
is dictated by the choice of simulation geometry. For Case-I and
Case-II we consider a plane circular slab 1-0 geometry. For these
two cases photons are uniformly ($r=r_{c}\sqrt{\xi}$,$\phi=2\pi\xi$)
and isotropically ($\mu_{inj}=\xi$,$\phi_{inj}=2\pi\xi$) injected
at the base for propagation in the upward hemisphere, where $r,\phi$
are the radial and azimuthal angle co-ordinate in cylindrical co-ordinate
system, and $\xi$ is a uniform random variate in the range 0 to 1. 

\begin{figure}
\includegraphics[scale=0.3]{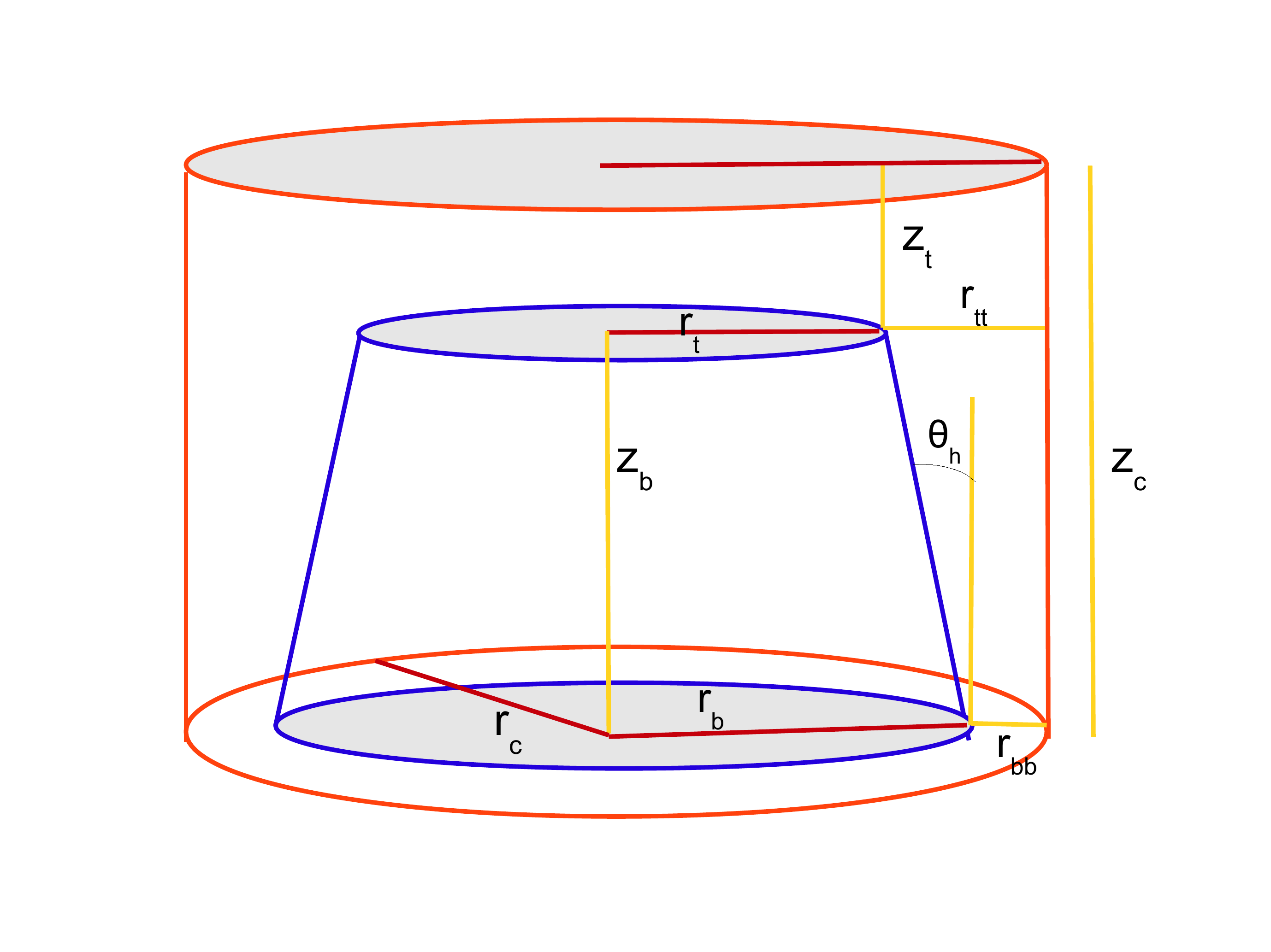}

\caption{\label{fig:f1}This figure depicts the geometry of the column for
the Case-III in Table--\ref{tab:4.1}. The structure in the blue color
is a cone cut at the top. The surface of the cone is defined such
that it lies at a fixed optical depth below the surface of the cylindrical
column.}
\end{figure}

For the cylindrical geometry (Case-III), we place the source plane
at a fixed optical depth $\tau_{T}=10^{-3}$ inside the boundaries
of the cylindrical column. The source plane can be approximated as
a truncated cone buried inside the cylindrical column (see  Fig--\ref{fig:f1}).
The scheme for selection of position co-ordinates of injection ($x_{inj}$,$y_{inj}$,$z_{inj}$)
at the source plane and the direction of injection ($\theta_{inj}$,$\phi_{inj}$)
for case III, is given in Appendix A. The injected photons diffuse
in the region bounded by the injection surface and the outer surface
of the cylindrical column.

\subsection{Free path of the photon\label{sec:4.4}}

The effective local mean free path in an inhomogeneous medium is given
by
\begin{align}
\bar{\lambda}(B^{'},\omega_{i},\mu_{i}) & =\dfrac{1}{n_{e}P_{tot}(B^{'},\omega_{i},\mu_{i})}\label{eq:4.14}
\end{align}
where $P_{tot}(B^{'},\omega_{i},\mu_{i})$ is the total line profile.
The line profile is computed as follows,
\begin{equation}
    P_{tot}=P_{abs}+P_{sc}
\end{equation}

\begin{equation}
\begin{split}
P_{g}(B^{'},\omega_{i},\mu_{i})&=\langle\sigma_{g}(B^{'},\omega_{i},\mu_{i})\rangle_{f_{e}(p)}\\
&=\int_{-\infty}^{\infty}\left(\sum_{n=n_{v}}^{n_{max}}\sigma_{g}^{n}(B^{'},\omega_{i},\mu_{i})\right)f_{e}(p)dp\label{eq:4.5}
\end{split}
\end{equation}
where $n_{e}$ is the local electron density, $\sigma_{abs}^{n}(B^{'},\omega_{i},\mu_{i})$
($g=abs)$ is the absorption and $\sigma_{sc}^{n}(B^{'},\omega_{i},\mu_{i})$($g=sc$)
is the total scattering cross-section from ground state to state $n$. Here,
$n_{v}=0$ for scattering and $n_{v}=1$ for absorption.
The incident
photon $\omega_{i}$ can excite the electron up a maximum Landau level
$n_{max}$ determined by $\omega_{i}\approx n_{max}B^{'}$. We use
fully relativistic absorption cross-section \citep{Harding_Daugherty_1991}  and scattering cross-section $\sigma_{sc}$ from \citet{Sina_1996}.
The scattering cross-section is spin and polarization
dependent, while we have used a polarization and spin averaged cross-section by summing over the final $(f)$ and averaging over the initial spin and polarization states.
The free path $\lambda,$ namely the distance travelled by a photon
before encountering an absorption or scattering, may then be obtained
from the probability distribution giving

\begin{equation}
\lambda=-\bar{\lambda}\text{ln}(1-\xi)=-\bar{\lambda}\text{ln}\xi\label{eq:4.16}
\end{equation}
 In the last equality ($1-\xi)$ has been replaced by $\xi$ as the
distribution of both are the same.

\subsection{Probability of absorption and scattering\label{sec:4.5}}
 A photon with energy $\omega_{i}$
and moving in the direction $\mu_{i}$ is either absorbed or scattered
by an electron after traveling a free path $\lambda$. Whether a scattering or an absorption will occur
is chosen based on the fractional probabilities

\[
R_{abs}(B^{'},\omega_{i},\mu_{i})=\dfrac{{\displaystyle \sum_{n=1}^{n_{max}}}\sigma_{abs}^{n}(B^{'},\omega_{i},\mu_{i})}{{\displaystyle \sum_{n=1}^{n_{max}}}\sigma_{abs}^{n}(B^{'},\omega_{i},\mu_{i})+{\displaystyle \sum_{n=0}^{n_{max}}}\sigma_{sc}^{n}(B^{'},\omega_{i},\mu_{i})}
\]

\[
R_{sc}(B^{'},\omega_{i},\mu_{i})=\dfrac{{\displaystyle \sum_{n=1}^{n_{max}}}\sigma_{sc}^{n}(B^{'},\omega_{i},\mu_{i})}{{\displaystyle \sum_{n=1}^{n_{max}}}\sigma_{abs}^{n}(B^{'},\omega_{i},\mu_{i})+{\displaystyle \sum_{n=0}^{n_{max}}}\sigma_{sc}^{n}(B^{'},\omega_{i},\mu_{i})}
\]
If the uniform variate $\xi<R_{abs}$ then absorption is selected,
otherwise scattering is selected.

\subsection{Momentum selection}

The selection of parallel momentum of the interacting electron is performed using a partial profile (see AH99 for more details),

\begin{equation}
P_{g}^{p}(B^{'},\omega_{i},\mu_{i},q_{i})=\int_{-\infty}^{q_{i}}\left(\sum_{n=n_{v}}^{n_{max}}\sigma_{g}^{n}(B^{'},\omega_{i},\mu_{i})\right)f_{e}(p)dp\label{eq:4.4}
\end{equation}
Here, $g=sc$ is for scattering and $g=abs$ is for the absorption cross-section.
The absorption cross-section $\sigma_{abs}$ and scattering cross-section
$\sigma_{sc}$ have similar values near resonance region and somewhat different values at the continuum. Run time computation
of partial profile $P_{sc}^{p}$ is extremely time consuming as compared
to $P_{abs}^{p}$ because of more complex analytic expression of $\sigma_{sc}$
containing infinite sums on virtual Landau states \citep{Sina_1996}.
Since the cross-section at resonance is orders of magnitude greater than the continuum, the contribution of near resonance
region dominates over the contribution of continuum in the
partial profiles. For the case of uniform field the momentum selection
is performed using pre-computed $P_{sc}^{p}(B^{'},\omega_{i},\mu_{i},q_{i})$
scattering partial profiles and for nonuniform magnetic field it is computed
runtime using $P_{abs}^{p}(B^{'},\omega_{i},\mu_{i},q_{i})$.

After selecting the momentum, Monte-Carlo modeling of the rest of the
processes including items 4 and 5 in the Sec. \ref{sub:radtran1} are
done in a manner similar to that in AH99.

\subsection{Numerical implementation \label{sec:4.6-1}}

We perform the Monte-Carlo simulations for two classes of magnetic
field a) uniform magnetic field and uniform density (Case-I of Table--\ref{tab:4.1}),
and b) for varying magnetic field and varying density (Case-II, Case-III
of Table--\ref{tab:4.1}). While the basic steps are the same for both
classes, the implementation differs considerably. It is noticeable
that nearly all PDFs have dependence on $(B^{'},\mu_{i})$. Three
reference frames are considered for performing the radiative trasfer, 
\begin{enumerate}
\item $Gb$ ($X^{G},Y^{G},Z^{G}$) with origin at the base of the mound
and z axis oriented along the magnetic axis) 
\item $Es$ $(X^{E},Y^{E},Z^{E})$ with origin at the point of scattering
or absorption, with its axis $X^{E},Y^{E},Z^{E}$ parallel to the
axis $X^{G},Y^{G},Z^{G}$ of the $Gb$
\item $Lb$ $(X^{L},Y^{L},Z^{L})$ attached to the point of scattering or
absorption with its $Z^{L}$ axis along the local magnetic field vector
and with $Y^{L},Z^{L}$ plane containing the $Z^{E}$
axis. 
\end{enumerate}
In case of varying magnetic field the direction of the magnetic field
vector changes from place to place so we have used all the three frames
($Gb$,$Es$,$Lb$) to deal with the situation. For the non-uniform magnetic
field case the radiative transfer is performed in three stages. 
\begin{description}
\item [\textbf{Stage-I:}] Injection is performed at a point $x_{i},y_{i},z_{i}$
on the source plane in $Gb$ frame in the direction $\mu_{i},\phi_{i}$.
\item [\textbf{Stage-II:}] The $\mu_{i},\phi_{i}$ are converted to $Lb$ frame
angles $\mu_{i}^{L},\phi_{i}^{L}$ using Eq--\ref{eq:4.7.1}.

\[
\Omega^{L}=T\Omega^{G}
\]
where $T$ is the transformation matrix and is given by \\
\begin{equation}
T=\left(\begin{array}{ccc}
\sin\phi_{B} & -\cos\phi_{B} & 0\\
\cos\theta_{B}\cos\phi_{B} & \cos\theta_{B}\sin\phi_{B} & -\sin\theta_{B}\\
\sin\theta_{B}\cos\phi_{B} & \sin\theta_{B}\sin\phi_{B} & \cos\theta_{B}
\end{array}\right)\label{eq:4.7.1}
\end{equation}
where $\theta_{B},\phi_{B}$ are the angles of magnetic field vector
(z axis of the $Lb$ frame) measured in the $Es$ or the $Gb$ frame. The transformation $Gb\rightarrow Lb$
facilitates the computation since the PDF of several processes do
not depend upon $\mu_{i}$ but depend upon the angle between the local
magnetic field vector and the photon propagation vector ($\mu_{i}^{L}$).
The computation of all the following steps related to scattering,
absorption and emission are preformed in the local frame $Lb$. 
\begin{enumerate}
\item Selection of the momentum ($p_{i}$). 
\item Selection of the free path $\lambda(\omega_{i},\mu_{i}^{L})$.
\item Selection between absorption or scattering, and for absorption we terminate the trajectory of the photon. 
\item Selection of scattering angle ($\mu_{f}^{L}$,$\phi_{f}^{L}$)
\item Computation of photon energy $\omega_{f}$ and final electron momentum $p_{f}$
\item Selection of the Landau level $n_{f}$ after excitation
\item Selection of the Landau level $n_{f}^{'}$ after de-excitation
\item Selection of emission angle $\mu_{t}^{L},\phi_{t}^{L}$ of the
transition photon.
\item Computation of the energy of the transition photon $\omega_{t}$
and the final momentum $p$ of the electron in state $n_{f}^{'}$ 
\end{enumerate}

\item [\textbf{Stage-III:}] The angle ($\mu_{f}^{L}$,$\phi_{f}^{L}$) of the scattered
photon and $\mu_{t}^{L},\phi_{t}^{L}$ of the transition photons are
transformed back to ($\mu_{f}$,$\phi_{f}$), ($\mu_{t},\phi_{t}$)
in $Gb$ frame using inverse transformation of Eq--\ref{eq:4.7.1}.

\end{description}
In case of uniform magnetic field, since the direction of the field
is the same everywhere, only one reference frame $Gb$ ($X^{G},Y^{G},Z^{G}$)
is sufficient for the radiative transfer. All the three stages of
radiative transfer stated above are performed in $Gb$ frame.

\subsubsection{Treating varying magnetic field and plasma parameters }

Since the magnetic field varies in the simulation region, the computation
of the free path is not very straightforward. For this reason we
have divided the simulation volume into cuboid cells within each of
which the local density and magnetic field are assumed constant at
the value evaluated at the center of the cell. The cell-to-cell radiative
transfer is performed by creating a new cell centered at the point
where the photon hits a cell edge. This method can in principle be used
for arbitrarily large gradients by making cell sizes small enough, but
with the practical limitation that for very small cells the computation
time increases significantly. We use this method only while treating
varying magnetic field and density in the simulation volume (Case-II
and Case-III of Table--\ref{tab:4.1}).

\subsubsection{Numerical computation of partial and total line profile functions\label{sub:4.7.2}}

Total line profiles $P(B^{'},\omega_{i},\mu_{i})$ are used to compute
the mean free path $\bar{\lambda}$ and partial profiles $P^{p}(B^{'},\omega_{i},\mu_{i},q_{i})$
are used in momentum selection.  The computation of the partial profiles
$P_{sc}^{p}(B^{'},\omega,\mu_{i},q_{i})$ and line profiles $P_{sc}(B^{'},\omega_{i},\mu_{i})$,
which involve the computation of scattering cross-section $\sigma_{sc}^{n}(B^{'},\omega_{i},\mu_{i})$,
is the most computationally expensive part of our code. To include
the natural line width $\Gamma^{n,s}(p)$ in the cross-section $\sigma_{sc}^{n}(B^{'},\omega_{i},\mu_{i})$,
we compute the transition rates $\Gamma^{n,s}$ in $p=0$ frame and
store these values on a grid of ($B^{'},n,s$) for $50$ equally spaced
values between ($B_{min},B_{max}$),$20$ values of $n\in[1,2...,20]$
 and two values of the spin ($s$). Where $B_{min}$,$B_{max}$ are the minimum
and maximum values of the magnetic field in the simulation region.
Later the computation of $\Gamma^{n,s}(p)$ which is needed in computation
of $\sigma_{sc}$ can be performed by doing Lorentz transformation
on stored values $\Gamma^{n,s}(p=0)$ with the transformation $\Gamma^{n,s}(p)=[m(1+2nB^{'})^{1/2}/E_{n}(p)]\Gamma^{n,s}(p=0)$.
The line profiles $P(B^{'},\omega_{i},\mu_{i})$ and partial profiles
$P^{p}(B^{'},\omega_{i},\mu_{i},q_{i})$ are computed and stored once
on a grid of ($B,\mu_{i},\omega_{i}$) and ($B,\mu_{i},\omega_{i},q_{i}$)
respectively and then utilized in the runtime. 

To store the line profile $P_{sc}(B^{'},\omega_{i},\mu_{i})$ on a
grid of $B^{'},\omega_{i},\mu_{i}$, we follow a scheme similar to that adopted by AH99, however the range of parameters and step sizes
are different in our case. We have taken 6 values from \textbf{$B^{'}=0.0199$}
to \textbf{$B^{'}=0.03$} on a equally spaced logarithmic interval
and 6 values from $B^{'}=0.039$ to $B^{'}=0.15$ on equally space logarithmic
interval, 100 equally spaced values for $\mu_{i}\in(-0.985,0.985)$.
Sampling of $\omega_{i}$ depends upon the value of $\mu_{i}$, which
is based on the fact that the resonant peaks are very sharp, occurring at $\omega_{i}\rightarrow\omega_{n}^{cut}$ (=$\sqrt{1+2nB'}-1)/\sin\theta_{i}^{'}$, for details see Eq 44 of AH99)
for $|\mu_{i}|<0.5$, but are relatively broad, occurring at $\omega_{i}\rightarrow\omega_{n}^{res}$
for $|\mu_{i}|>0.5$. The estimate of the width of the resonance peak
can be approximated by Doppler width $w_{D}=B'\sqrt{2mT_e}$. In case of
$|\mu_{i}|<0.5$ two different grids on $\omega_{i}$ are made: 1)
a uniform grid $U_{\omega_{i}}$on $\omega_{i}$ and 2) a binary spacing
grid $H_{\omega_{i}}$ closer to the each peak $\omega_{n}^{cut}$.
In case of a binary spacing grid $H_{\omega_{i}}$ the sampling begins
at the point $\omega_{n}^{cut}-w_{D}$ on the left and the point $\omega_{n}^{cut}+w_{D}$
on the right and approaches until $|1-\omega_{i} / \omega_{n}^{cut}|<10^{-3}$
in step lengths that are progressively halved. Finally both the grids
$U_{\omega_{i}}$ and $H_{\omega_{i}}$ are merged to form a final
grid $F_{\omega_{i}}$. For the case of $|\mu_{i}|>0.5$ equal interval
sampling is employed throughout and $\omega_{i}$ values are saved
on a uniform grid $U_{\omega_{i}}$. 

Storing the partial profiles $P_{sc}^{p}(B,\omega_{i},\mu_{i},q_{i})$
on a grid needs to save one more parameter $p_{i}$ along with $B,\omega_{i},\mu_{i}$.
The sampling scheme for $B,\omega_{i},\mu_{i}$ is the same for $P_{sc}(B^{'},\omega_{i},\mu_{i})$. Placing the sampling points
for parallel momentum $p_{i}$ needs extra care since the width, height
and position of the resonance peaks in $p_{i}$ space are highly variable
depending upon the values of $B,\omega_{i},\mu_{i}$. For each value
of $n$, zero, or one ($p_{n0}^{s}$) or two ($p_{n+}^{s},p_{n-}^{s})$
solutions of the equation $\omega_{i}^{'}=\omega_{n}^{res}(\mu_{i}^{'})$
are possible (AH99). So all the possible solutions from $n=1$ to
$n=10$ which lie in the range ($-50T_{e},50T_{e}$) are stored. Sampling
of $p_{i}$ is performed on two grids, a uniform grid $U_{p_{i}}$
in $p_{i}$ and a binary spacing grid $H_{p_{i}}$ near each solution
in $p_{i}$. Finally after merging these two grids and sorting in
$p_{i}$, a final grid is obtained on which partial profiles $P_{sc}^{p}(B^{'},\omega_{i},\mu_{i},q_{i})$
are stored. 

The mean free path $\bar{\lambda}=1/(n_{e}(P_{sc}+P_{abs}))$ is selected
by using the line profiles for both scattering and absorption where
$P_{abs}(B^{'},\omega_{i},\mu_{i})$ is computed runtime and $P_{sc}(B^{'},\omega_{i},\mu_{i})$
is obtained by interpolation of the
already stored values of $P_{sc}(B^{'},\omega_{i},\mu_{i})$. In case
of varying magnetic field, for both absorption and scattering, we
use $P_{abs}^{p}(B^{'},\omega_{i},\mu_{i},q_{i})$ for momentum selection
since the interpolation of partial profiles $P_{sc}^{p}(B^{'},\omega_{i},\mu_{i},q_{i})$
across different $B^{'}$ turn out to be not accurate (due to the
extreme sensitivity of the position and features of the resonance
peaks in momentum space on the parameters $B^{'},\omega_{i},\mu_{i},p_{i}$).
We compute partial profile $P_{abs}^{p}(B^{'},\omega_{i},\mu_{i},q_{i})$
runtime on exact values of $B^{'},\omega_{i},\mu_{i}$ to select the
momentum. The use of absorption partial profile $P_{abs}^{p}(B^{'},\omega_{i},\mu_{i},q_{i})$
for momentum selection of scattering process is justified near the resonance energy since both
the cross-sections have similar numerical values in that region.

\subsubsection{Interpolation in the table of stored values of line profile functions
$P_{sc}(B^{'},\omega_{i},\mu_{i})$ }

The line profiles are stored on a grid of magnetic field, cosine of
angle and energy ($(B_{st}^{'}(i),i=1...12)$, $(\mu_{st}(j),j=1...100)$,
$(\omega_{st}(k),k=1...k_{max})$) with 12 values of the magnetic
field, 100 equally spaced values of $\mu_{i}$ and strategically placed
$k_{max}$ energy values as discussed in Sec.\ref{sub:4.7.2}. An
interpolation scheme is needed to compute the line profile $P_{sc}(B^{'},\omega_{i},\mu_{i})$
at any other value of magnetic field $B_{com}^{'}$, energy $\omega_{com}$
and angle $\mu_{com}$. First we select the two values of magnetic
field $B_{st}^{'}(i)$, $B_{st}^{'}(i+1)$ on a grid between which
the value $B_{com}^{'}$ lies. Next we locate a $\mu_{ns}$ value
which is nearest to $\mu_{com}$ on the grid $\mu_{st}$. We choose
to perform interpolation for this value $\mu_{ns}$. We have now two
line profiles $P_{sc}(B_{st}^{'}(i),$ $\mu_{ns},$ $(\omega_{st}(k),k=1,k_{max}))$,
$P_{sc}(B_{st}^{'}(i+1),$ $\mu_{ns},$ $(\omega_{st}(k),k=1,k_{max}))$
and interpolation is to be performed on these line profiles. The line
profiles are resonant at energies ($\omega_{1}^{res}$, $\omega_{2}^{res}...)$
and the resonance energies $\omega_{n}^{res}$ scale with the strength
of the magnetic field. The line profiles, too, scale in energy space
in proportion to the resonance energies. We utilize this scaling behavior
to interpolate the line profile. We first find the two scaled energy
values $\omega_{l}(\omega_{com},$ $B_{com}^{'},$ $B_{st}^{'}(i))$
and $\omega_{f}$$(\omega_{com},$ $B_{com}^{'}$, $B_{st}^{'}(i+1))$
which provide the energy markers on the two values $P_{sc}$$(B_{st}^{'}(i),$
$\mu_{ns}$, $\omega_{l})$ and $P_{sc}(B_{st}^{'}(i+1),$ $\mu_{ns}$,
$\omega_{f})$ between which the interpolation will be performed to
derive $P_{sc}$$(B_{com}^{'},$ $\mu_{ns},$ $\omega_{com})$. The
values of $\omega_{l}$ and $\omega_{f}$ are derived as follows.
We compute the first 10 resonance energies $\omega_{n}^{res}(B^{'},$
$\mu_{ns})$ values for three magnetic fields $B_{st}^{'}(i)$, $B_{com}^{'}$,
$B_{st}^{'}(i+1)$, i.e three lists of resonance energies are created
for three magnetic fields: $T_{f}=[\omega_{n}^{res}($$B_{st}^{'}(i+1)$,
$\mu_{ns}),$ $n=1,..10],$ $T_{m}=[\omega_{n}^{res}($$B_{com}^{'},$
$\mu_{ns}),$ $n=1,..10]$ , $T_{l}=[\omega_{n}^{res}($\textbf{$B_{st}^{'}(i),$
$\mu_{ns},$ $n=1,..10]$}. Now a second order polynomial fit is
performed on the points $T_{f}$ versus $T_{m}$ ( say with a fitting
polynomial $f_{f}(\omega_{i})$) and a similar one on $T_{l}$ versus
$T_{m}$ with another fitting polynomial $f_{l}(\omega_{i})$. From
these, $\omega_{f}$ and $\omega_{l}$ are derived as follows: 

\[
\begin{array}{c}
\omega_{f}=f_{f}(\omega_{i}=\omega_{com})\\
\omega_{l}=f_{l}(\omega_{i}=\omega_{com})
\end{array}
\]
 One should note that $\omega_{f}$ and $\omega_{l}$ are computed
using the resonance energies $\omega_{n}^{res}(B^{'})$ which increase
with increasing magnetic field, hence $\omega_{l}<\omega_{com}<\omega_{f}$.
We have stored the $P_{sc}$ values for $\omega_{i}\in(0.1-200)$ \text{KeV}
so it is possible that at energies at the boundary close to 1 KeV and
200 KeV the values of $\omega_{f}$, $\omega_{l}$ may go out of bound
(i.e.$\omega_{l}<0.1$ \text{KeV} or $\omega_{f}>200 $ \text{KeV}) hence
we restrict our simulation to ($1-150 \text{) KeV}$ and at times we
simulate only in the continuum range ($1-100$) \text{KeV}. 

In each of the two arrays $P_{sc}(B_{st}^{'}(i)$, $\mu_{ns},$ ($\omega_{st}(k),k=1,k_{max}))$
and $P_{sc}(B_{st}^{'}(i+1)$, $\mu_{ns},$ ($\omega_{st}(k),k=1,k_{max}))$
we then linearly interpolate on $\omega$ to derive $P_{sc}(B_{st}^{'}(i),$$\mu_{ns},$ $\omega_{l}$) and $P_{sc}(B_{st}^{'}(i+1),$ $\mu_{ns},$
$\omega_{f})$ respectively. Finally, a linear interpolation on $B^{'}$
is carried out between these two values to obtain our desired quantity
$P_{sc}(B_{com}^{'},$ $\mu_{com},$ $\omega_{com}).$

\subsection{Tests of the Monte Carlo code }

We performed several checks and tests of our Monte-Carlo code and
two of them are presented here.

\subsubsection{Escape probability\label{sub:Escape-probability}}

\begin{figure}
\includegraphics[scale=0.6]{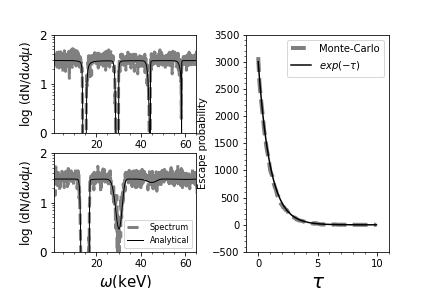}

\caption{\label{fig:f3}{The simulation is performed
for 1-0 Slab }($r_{c}=1\text{ Km}$, $h_{c}=100\text{ cm}$) for a
uniform magnetic field $B^{'}=0.03$, $T_{e}=5 \text{ KeV}$, $\tau_{T}(\mu_{i}=1)=10^{-3}$ {.Left
panel: Comparison of the Monte-Carlo and analytical estimates of the
pencil beam spectra. The dashed line shows the Monte Carlo results
and the solid line the analytical expectation. The top left panel is
for $\mu_{i}=0.1$ and the bottom left panel is for $\mu_{i}=0.5$. Right
panel: The distribution of the optical depth $\tau$ corresponding
to the free paths generated for different scatterings. The dashed line shows the distribution generated in the Monte-Carlo simulation. The
solid line represents the probability function $\exp(-\tau)$ scaled
to the total number of scatterings.}}
\end{figure}

First, we performed a test to confirm that the escape probability
is modelled properly in our code. The simulation was performed for
a thin slab 1-0 ($r_{c}=1\text{ Km}$, $h_{c}=100\text{ cm}$) for
a uniform magnetic field $B^{'}=0.03$, $T_{e}=5 \text{ KeV}$, $\tau_{T}(\mu_{i}=1)=10^{-3}$.
The escape probability of a photon traversing an optical depth $\delta\tau(\omega_{i},\mu_{i})$
is given by $\exp(-\delta\tau(\omega_{i},\mu_{i}))$. We perform our
simulations in real space so this corresponds to,
\[
\delta\tau(\omega_{i},\mu_{i})=\dfrac{\delta\lambda}{\bar{\lambda}}=n_{e}\left\langle \sigma_{sc}(\omega_{i},\mu_{i})\right\rangle _{f_{e}(p)}\delta\lambda
\]
 we generate the free paths $\delta\lambda$ in our simulation and
store the corresponding $\delta\tau$ $(\omega_{i},\mu_{i})$ values,
the distribution of which is plotted in right panel of  Fig--\ref{fig:f3}.
The thick gray dashed line is for the actual distribution of $\delta\tau$ produced
by the Monte-Carlo simulations and the solid line represents the evaluation of
the function $\exp(-\delta\tau(\omega_{i},\mu_{i}))$ . The two results
match well, indicating that the escape probability is modelled accurately
in our code. 

\subsubsection{Pencil beam injection test}

Next, we performed a pencil beam injection test to confirm that the
intensity removed from the pencil beam is in accordance with the analytical
expectation from radiative transfer. The input parameters for this
case is same as set in Sec.\ref{sub:Escape-probability}. We injected
a beam of photons in a very small angle bin $\mu_{i}\pm\delta\mu_{i}$
with the energy in full continuum range (1-100 KeV). The escaping
photons were collected in the same angle range $\mu_{i}\pm\delta\mu_{i}$\@.
During the propagation many photons are scattered out of the beam.
The output spectrum of the photons remaining in the pencil beam is
expected to be given by just the optical depth profile as 

\begin{equation}
I(\omega_{i},\mu_{i})=I_{0}\exp(-\tau(\omega_{i},\mu_{i}))\label{eq:5.1}
\end{equation}
 where $I_{0}$ is the intensity of the injected continuum. The optical
depth $\tau$ in the direction $\mu_{i}$ is given by

\begin{equation}
\tau(\omega_{i},\mu_{i})=n_{e}\langle\sigma_{sc}(\omega_{i},\mu_{i})\rangle_{f_{e}(p)}\dfrac{h}{\mu_{i}}\label{eq:5.1.1}
\end{equation}
where $h$ is vertical height of the slab, $\mu_{i}=\cos\theta_{i}$,
$\theta_{i}$is the angle between perpendicular to the slab and viewing
direction and other symbols have their usual meaning.

The left panels of  Fig--\ref{fig:f3} show the comparison between the expectation from Eq--\ref{eq:5.1} with that of a
Monte Carlo run for constant magnetic field $B^{'}=0.03$, $\tau_{T}=10^{-3}$
and electron temperature $T_{e}=5$ KeV. The top left panel is for $\mu_{i}=0.1$ and the bottom left panel is for $\mu_{i}=0.5$.
The solid black lines represent the analytic estimate of the photon counts
corresponding to the specific Intensity $I(\omega_{i},\mu_{i})$ derived
from Eq--\ref{eq:5.1} and the gray dashed lines represent the spectra from
Monte-Carlo Simulations. They both agree well with each other. This test confirms that photon removal from the beam
due to scattering is accurately modelled in our code.

\section{Results} \label{sec4}

In this section we present the phase averaged spectra
in four angle bins. We will use the following abbreviated notation
repeatedly for these 4 angle bins: $\mu_1$ for the angle bin
$0.0<\mu_{i}\leq0.25$, $\mu_2$ for $0.25<\mu_{i}\leq0.5$, $\mu_3$
for a $0.5<\mu_{i}\leq0.75$, and $\mu_4$ for angle bin $0.75<\mu_{i}\leq1.00$.

\paragraph*{Redshift: }

All the spectra are gravitational redshift corrected. The gravitational
redshift factor near the neutron star surface is given by $\zeta =\sqrt{1-2GM_{*}/r}$,
with mass of the neutron star $M_{*}=1.4M_{\odot}$,$r\approx R_{*}=10\text{ Km}$,
gives the value $\zeta \approx0.77$.

\paragraph*{Continuum Model:}

The continuum model which we have used in all the spectra is HCUT
(Eq-- \ref{eq:6.2.3.1}), for parameters $\Gamma=0.91$, $\omega_{cut}=25.5 \text{ KeV}$ and
$\omega_{fold}=9.0 \text{ KeV}$.

\paragraph*{Light bending:}
The photons are re-distributed in angle due to strong light bending
effects  near the stellar surface.  We included this in computing the final observed spectra, as has also been done in  \citet{Nishimura_2019}. While computing the light bending two frames
are considered: 1) a column centric frame $Gb$($X^{G},Y^{G},Z^{G}$)
(as defined in Sec.\ref{sec:4.6-1}, 2) a star centric co-ordinate
system $St$($X^{S},Y^{S},Z^{S}$) which has its origin at the center
of the star and $z$ axis along the spin axis of the neutron star. We
will mention the ``spectra without light bending'' and ``spectra
with light bending'' frequently. The spectra without light bending
are produced in $Gb$ frame and light bending effects are not incorporated,
in this case $\mu$ is measured w.r.t the $Z^{G}$ axis of the $Gb$
frame. The spectra with light bending are produced in the $St$ frame,
computed after the inclusion of light bending. In this case $\mu$ is
measured from the $Z^{S}$ axis of the $St$ frame. The angle between
the spin axis of the neutron star and its magnetic axis is denoted
by $\theta_{B}$.  (see Appendix B for our implementation
of light bending). To include the effect from  both  poles of the NS, we have assumed that the magnetic axis of the second pole is at an angle
($\pi+\theta_B)$ (diametrically opposite location) with the spin axis of the NS. To compute the phase average spectrum we have taken the average of the spectra produced by the two poles.

\subsection{Monte Carlo simulations for Slab 1-0 with uniform magnetic fields }

The emission geometry for this case is taken to be slab $1-0$ (illuminated
from below, $r_{c}=1\text{ Km}$, $h_{c}=100\text{ cm}$) for uniform
field \textbf{$B^{'}=0.03$}, optical depth $\tau_{r}(\mu=1)=10^{-3}$.
Photons are injected uniformly and isotropically at the base of the
slab. It is assumed that if a photon hits the base of the slab then
it is absorbed by the base.

The phase averaged redshifted spectra for flat continuum are shown
in the left panel of the  Fig--\ref{fig:f5}. These clearly display
the strong angle and energy dependence of the emergent intensity.
At small angles $\mu\sim1$ the cyclotron features are shallow and
broad and at angles near $90^{\circ}$($\mu\rightarrow0$) they
are deeper and sharper. The first harmonic (fundamental) has a complex
shape in comparison to the second or higher harmonics. The fundamental
is shallower than the second harmonic and displays emission wings contributed
mainly by transition photons. In the low angle bins $\mu_3$ and
$\mu_4$, the emissions wings are more prominent. This results from
the transition photons being heavily scattered in high angle bins
due to large optical depth, and being thus redistributed to lower
angle bins before escape. The filling in of the fundamental due to
energy redistribution in multiple scatterings is therefore further
exacerbated by added transition photons at the low angle bins. Observations
also support that the fundamental is of a complex shape. In the source
4U0115+63 \citet{Heindl_et_al_2004} found that the fundamental is modeled
poorly by a Gaussian because of its very complex shape. Our results
for a flat continuum are similar to those of SC17 for the slab geometry.

\subsubsection{Effect of continuum}

\begin{figure}
\includegraphics[scale=0.6]{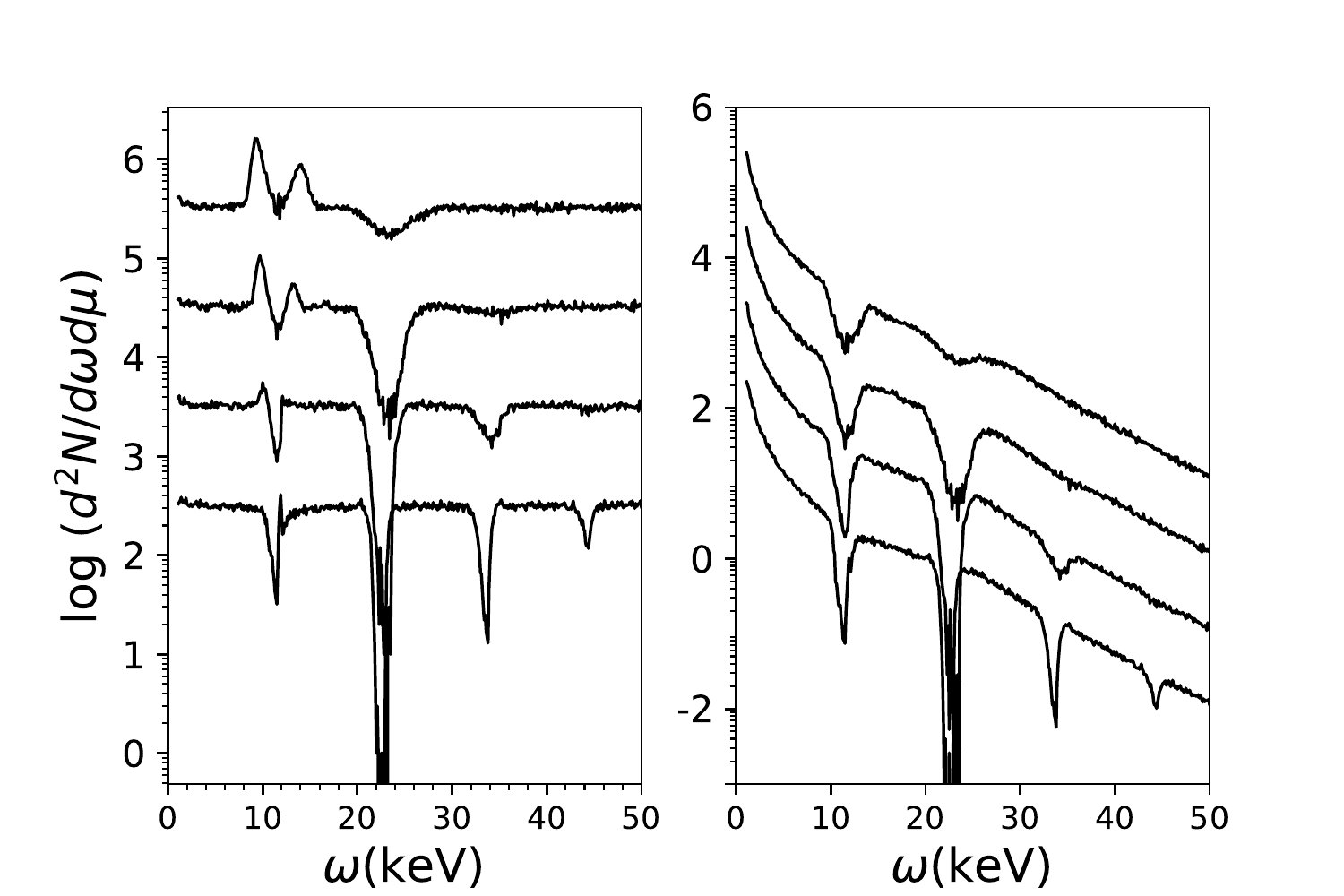}

\caption{\label{fig:f5} {The phase averaged redshifted
spectra, without light bending, for slab 1-0 geometry with uniform
field $B^{'}=0.03$, uniform electron temperature $T_{e}=5\text{ KeV}$
and optical depth $\tau_{T}=10^{-3}$. The left panel shows the spectra
for flat continuum $f_{p}(\omega)=1$ and the right panel shows the
spectra for HCUT continuum with $\Gamma=0.91$, $E_\text{cut}=25.5\text{ KeV}$,
$E_\text{fold}=9.0\text{ KeV}$. The spectra are plotted for four different
viewing bins $\mu_1$, $\mu_2$, $\mu_3$, $\mu_4$ from bottom to top.}}
\end{figure}

The right panel of the  Fig--\ref{fig:f5} shows the phase averaged
redshifted spectra for a \texttt{highecut} continuum model. We can see that the continuum significantly modifies the cyclotron
lines features. 
For the \texttt{highecut} continuum, the prominent emission wings
disappear, as the number of photons at higher energies which create transition photons at the fundamental are now much less. 
We find that the depth of the fundamental feature at low angles ($\mu_4$) has increased. So, with the \texttt{highecut} continuum, the fundamental is found to be deeper than the first harmonic. These features are shared by the spectra reported in Schon07
although their plasma parameters and the continuum model are slightly
different.

\subsubsection{Effect of temperature}

\begin{figure}
\includegraphics[scale=0.55]{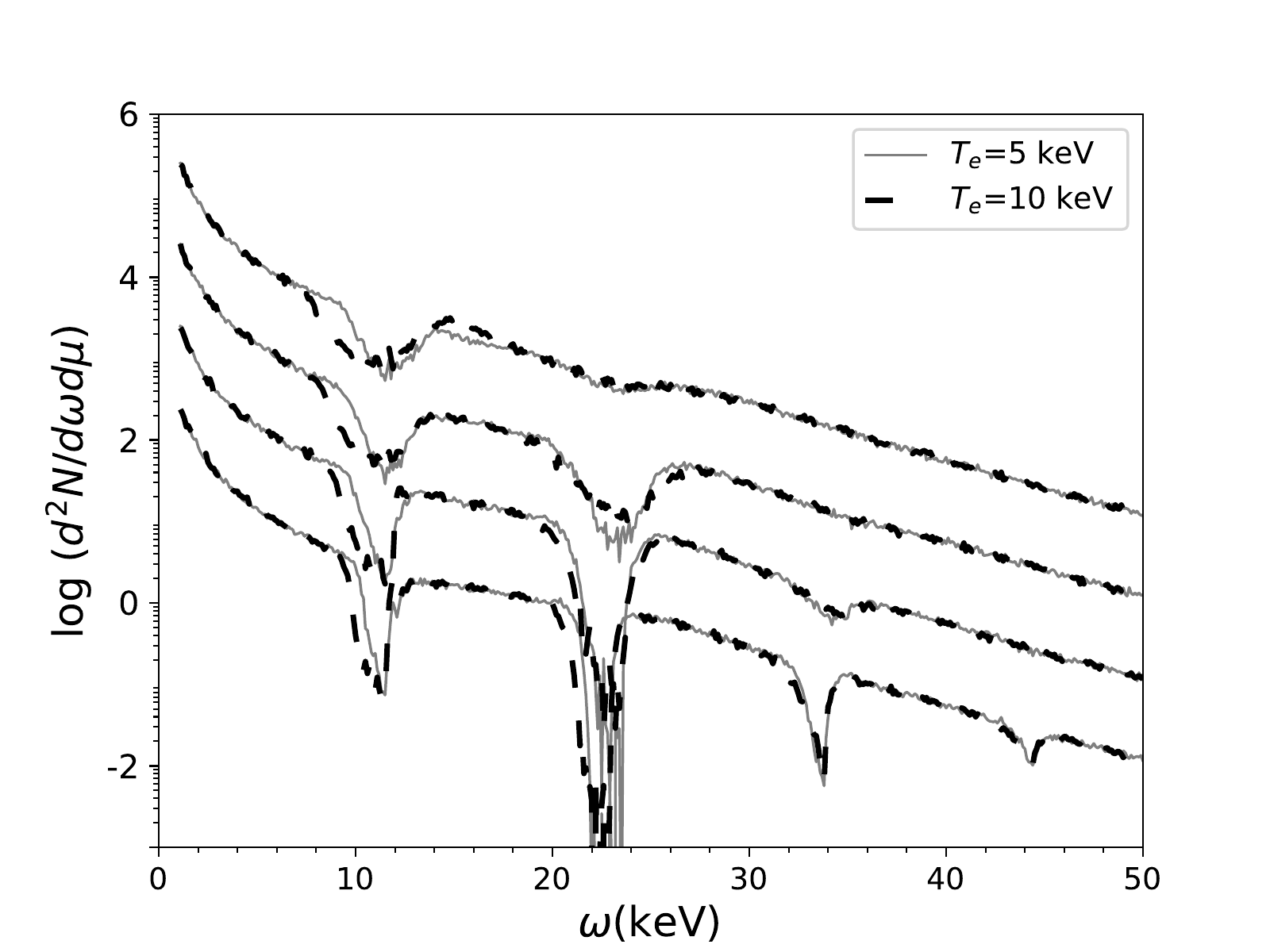}

\caption{\label{fig:f6} {The phase averaged redshifted
spectra, without light bending, for HCUT continuum and for the same
parameters as in  Fig--\ref{fig:f5} except for two different temperatures
$T_{e}=5\text{ KeV}$ (solid gray lines) and $T_{e}=10\text{ KeV}$ (dashed black line.)}}
\end{figure}

 Fig--\ref{fig:f6} shows the effect of electron temperature on the
cyclotron spectra. We have produced spectra for two different electron temperatures,
$5\text{ KeV}$ and $10\text{ KeV}$. For T$_e$=$10\text{ KeV}$, the cyclotron lines are found to be broader and shallower. At high
angle bins $\mu_1$, $\mu_2$, the broadening is also found to be asymmetric and skewed towards lower energies. This results from the
asymmetry in the peaks of the line profiles $P_{sc}(\omega_{i},\mu_{i})$
due to the relativistic cut-off in energy ($\omega_{n}^{cut}$) in
higher angles bins. 

\subsubsection{Effect of light bending}
\begin{figure}
\includegraphics[scale=0.55]{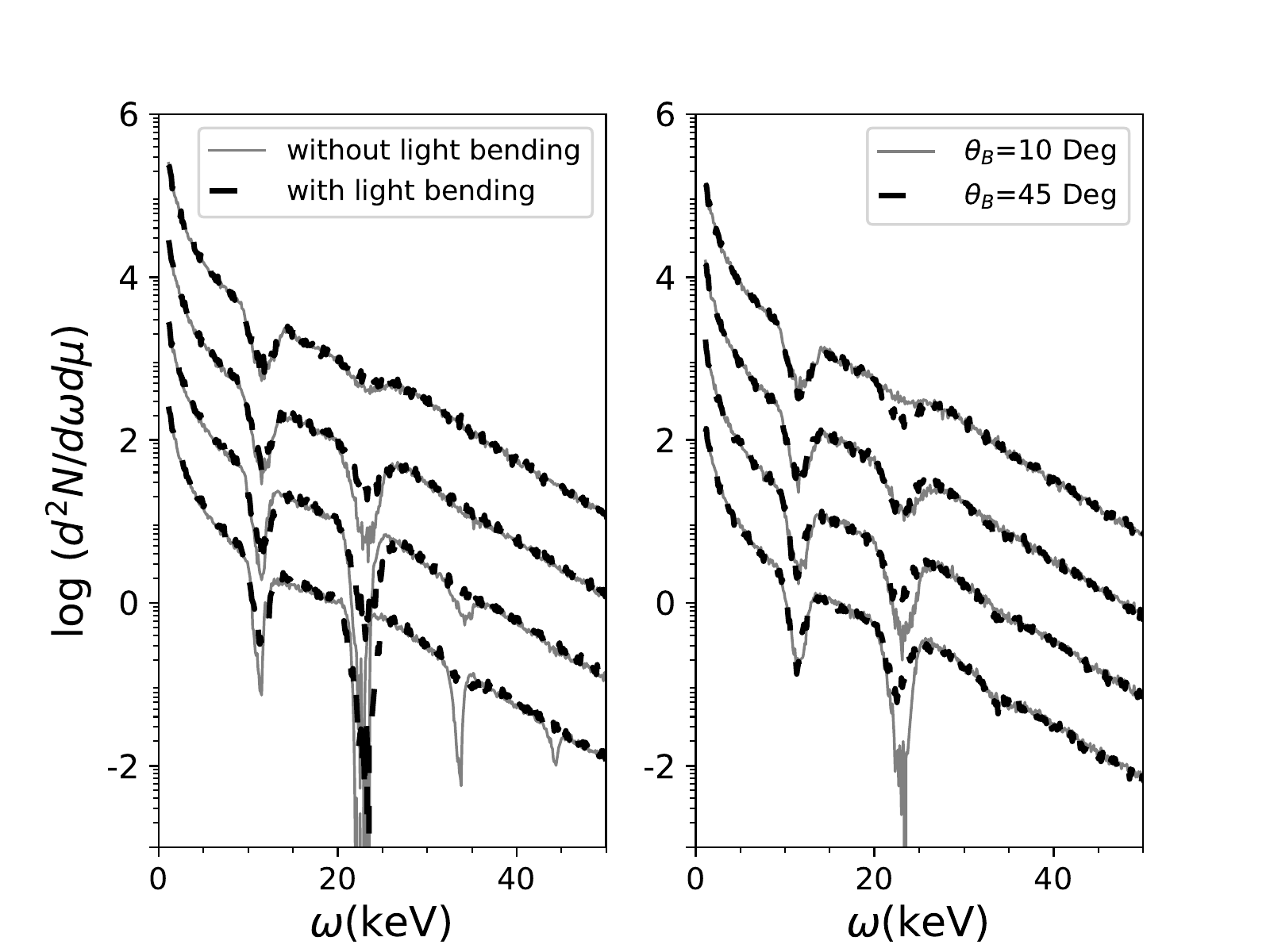}

\caption{\label{fig:f7} {Left panel: comparison of
spin phase averaged redshifted spectra with (dashed black)
and without light bending (solid gray), for $\theta_{B}=10^{\circ}$. Right panel: Spin phase averaged redshifted
spectra with light bending for two different angles $\theta_{B}=10^{\circ}$ (solid gray), and 45$^{\circ}$ (dashed black),
between spin axis and magnetic axis. The input parameters are the same
as in  Fig--\ref{fig:f5}.} The spectra are plotted for four different
viewing bins $\mu_1$, $\mu_2$, $\mu_3$, $\mu_4$ from bottom to top.}
\end{figure}

\begin{figure}
\includegraphics[scale=0.55]{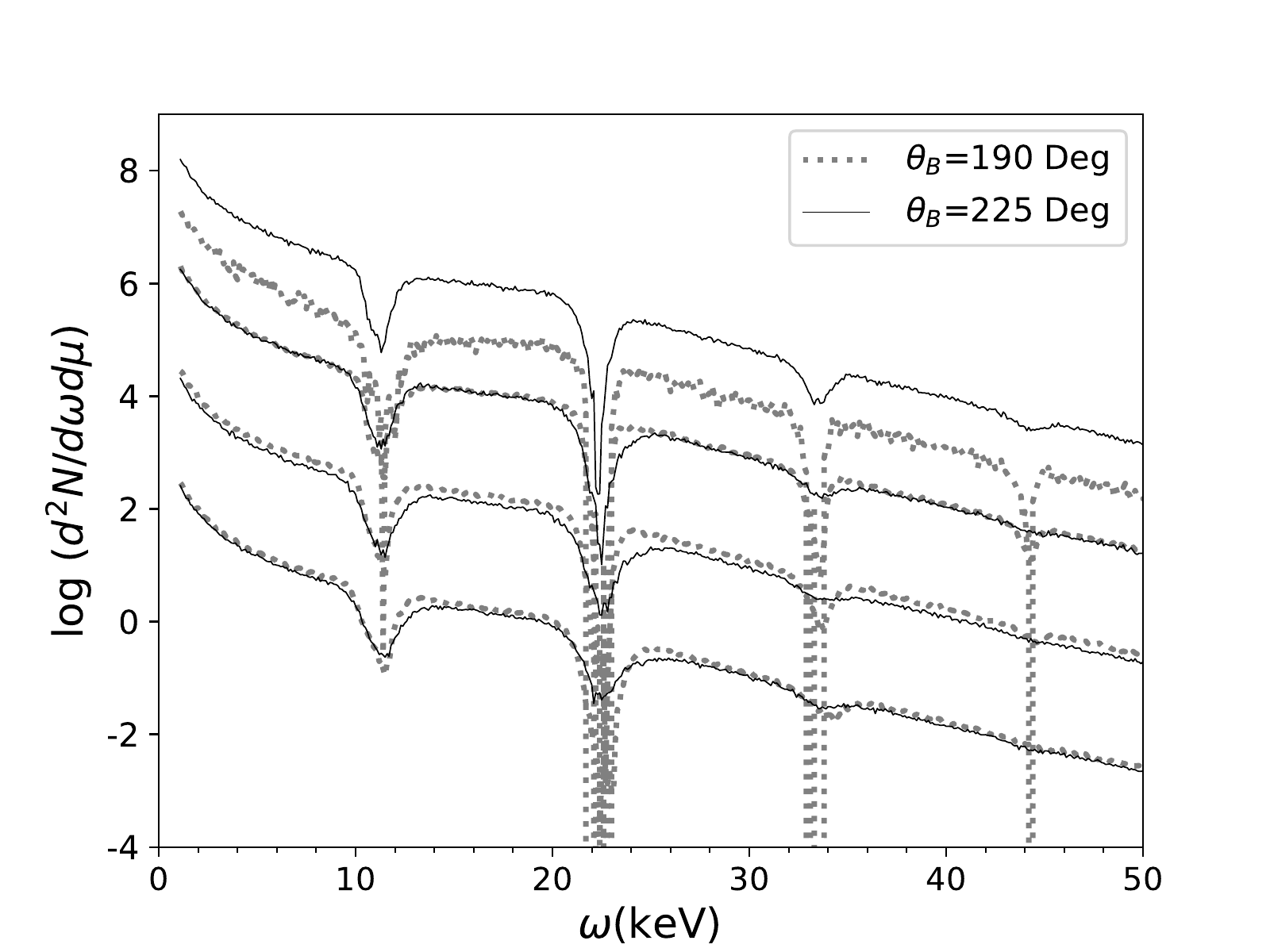}

\caption{\label{fig:slab_lb_pole2} {Phase averaged redshifted spectra with light bending for magnetic inclination  $\theta_{B}=190^{\circ}$ (broken gray), and $\theta_{B}=225^{\circ}$ (solid black). The input parameters are the same
as in  Fig--\ref{fig:f5}.}}
\end{figure}

\begin{figure}
\includegraphics[scale=0.55]{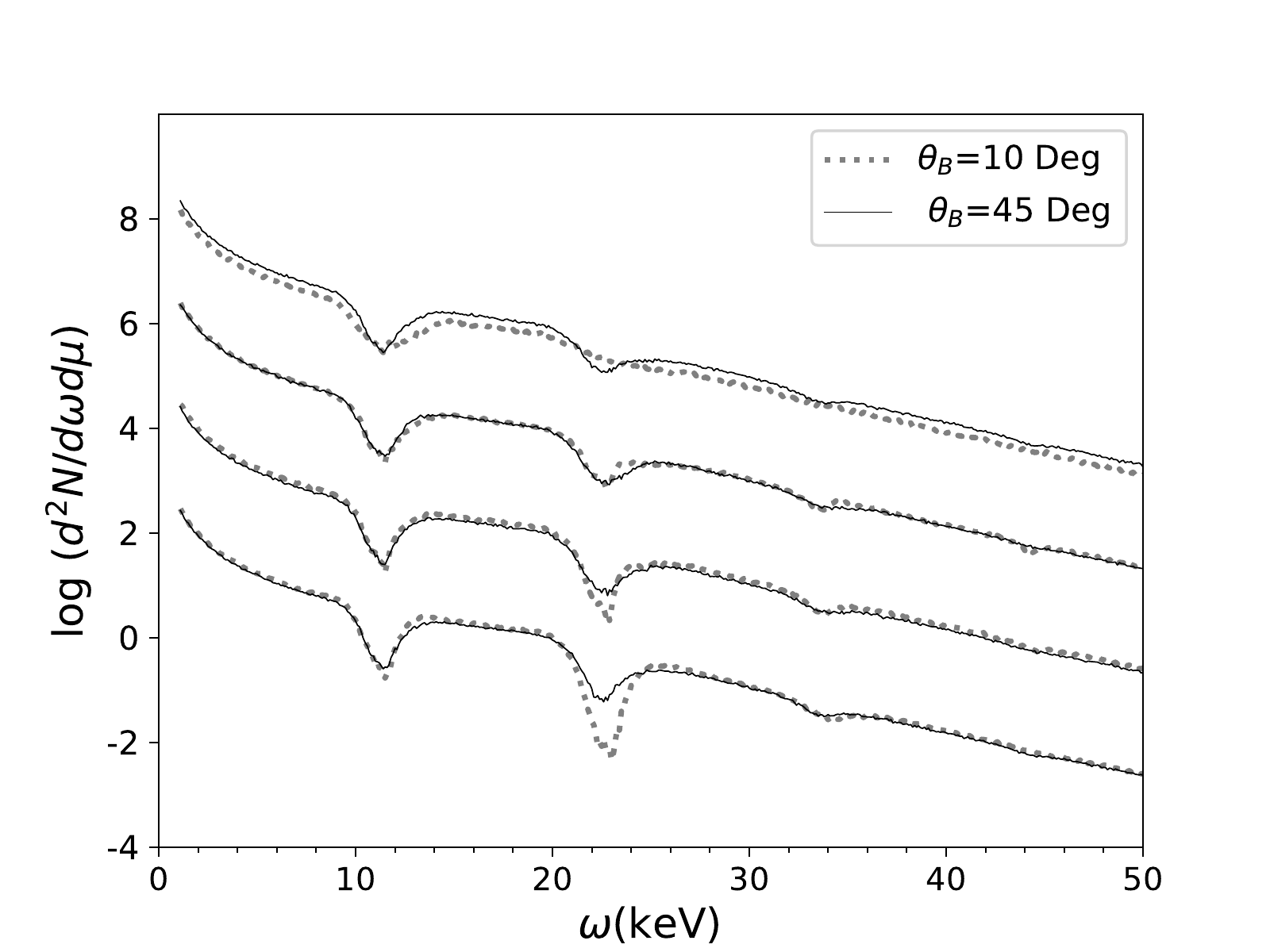}

\caption{\label{fig:slab_lb_bothpole} {Phase averaged redshifted spectra with light bending including both poles of the NS, for magnetic inclination $\theta_{B}=10^{\circ}$ (broken gray), and $\theta_{B}=45^{\circ}$ (solid black). The input parameters are the same
as in  Fig--\ref{fig:f5}.}}
\end{figure}

 The left panel of Fig--\ref{fig:f7} shows the CRSF without (solid gray lines) and with light bending (black dashed lines) assuming the angle between the spin axis and magnetic axis, $\theta_{B}=10^{\circ}$.
The right panel in this figure contains two plots with light
bending, one for $\theta_{B}=10^{\circ}$ (gray solid line), and the other for $\theta_{B}=45^{\circ}$(black dashed lines). Note that $\mu$ in
the case of spectra without light bending and that for the case with light bending
are measured from two different axis,
as already mentioned in the beginning of the Sec.\ref{sec4}. 
 The spectra with light bending represent the spin-phase averaged spectra which is normally observed from cyclotron line sources. Spectrum for any value of $\mu$ measured from the spin axis clearly includes the contribution from a range of $\mu$ values with respect to the magnetic axis. 
Thus the spectra with light bending involve an average over those without
light bending as can be seen in the left panel of  Fig--\ref{fig:f7}.
This tends to wash out sharp, highly angle-dependent features in
the spectrum as is evident from the figure - sharp, deep features
become shallow and wide, and some features, for example, harmonics
above the second, disappear altogether.
 The spectra at different angles look even more similar if the inclination angle between the spin axis and the magnetic axis is increased, as seen in the right panel of  Fig--\ref{fig:f7} where an inclination angle of $\theta_{B}=45^{\circ}$ is assumed. In Fig-\ref{fig:slab_lb_pole2} we have shown the phase averaged spectra from a slab for angles $\theta_{B}=190^{\circ}$ and $\theta_{B}=225^{\circ}$ with the spin axis. Fig-\ref{fig:slab_lb_bothpole}
shows the spin phase averaged redshifted spectra with light bending including both the poles of the NS, for $\theta_{B}=10^{\circ}$ and $\theta_{B}=45^{\circ}$.

 The main consequence of light bending is that if the spectrum originates close to the neutron star surface where light bending is strong, the difference between spectra observed at different angles tends to diminish considerably (Fig-\ref{fig:slab_lb_bothpole}).
This also drastically reduces the spin phase dependence of the spectrum, as also pointed out by MB12. 
This suggests that strongly phase-dependent spectra should either originate at large heights from the stellar surface or have contributions from physically different regions, such as asymmetric opposite poles (MB12).
 
Observationally, both significant (e.g. GX 301--2 \citet{Heindl_et_al_2004}) and small  (e.g.
V0332+53 \citet{Pottschmidt_et_al_2005}) dependence of the CRSF on spin phase have been observed. The reason for such diversity
is yet to be clearly understood, but is likely to lie in the diversity
of emission geometry.

\subsection{Accretion mound with accretion induced distorted magnetic fields}
Cyclotron lines bear the signature of the distortion of the magnetic
fields. In an accretion mound the magnetic field is severely distorted which could be reflected in the observed CRSFs.

\begin{figure}
\includegraphics[scale=0.6]{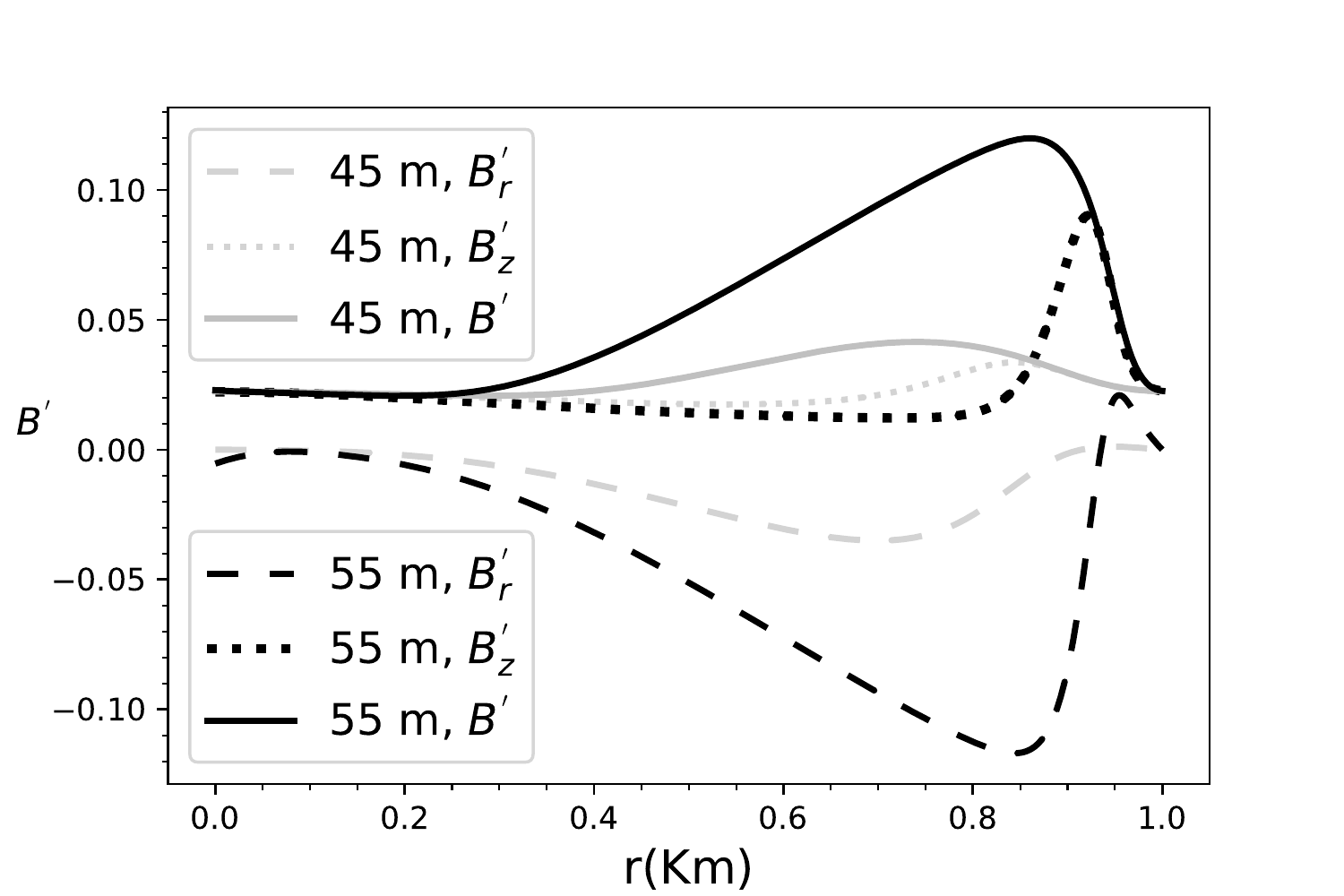}

\caption{\label{fig:f9}Radial dependence of the magnetic field components
at the top of the two mounds, of height 45m and 55m respectively.}
 
\end{figure}

\begin{figure}
\includegraphics[scale=0.6
]{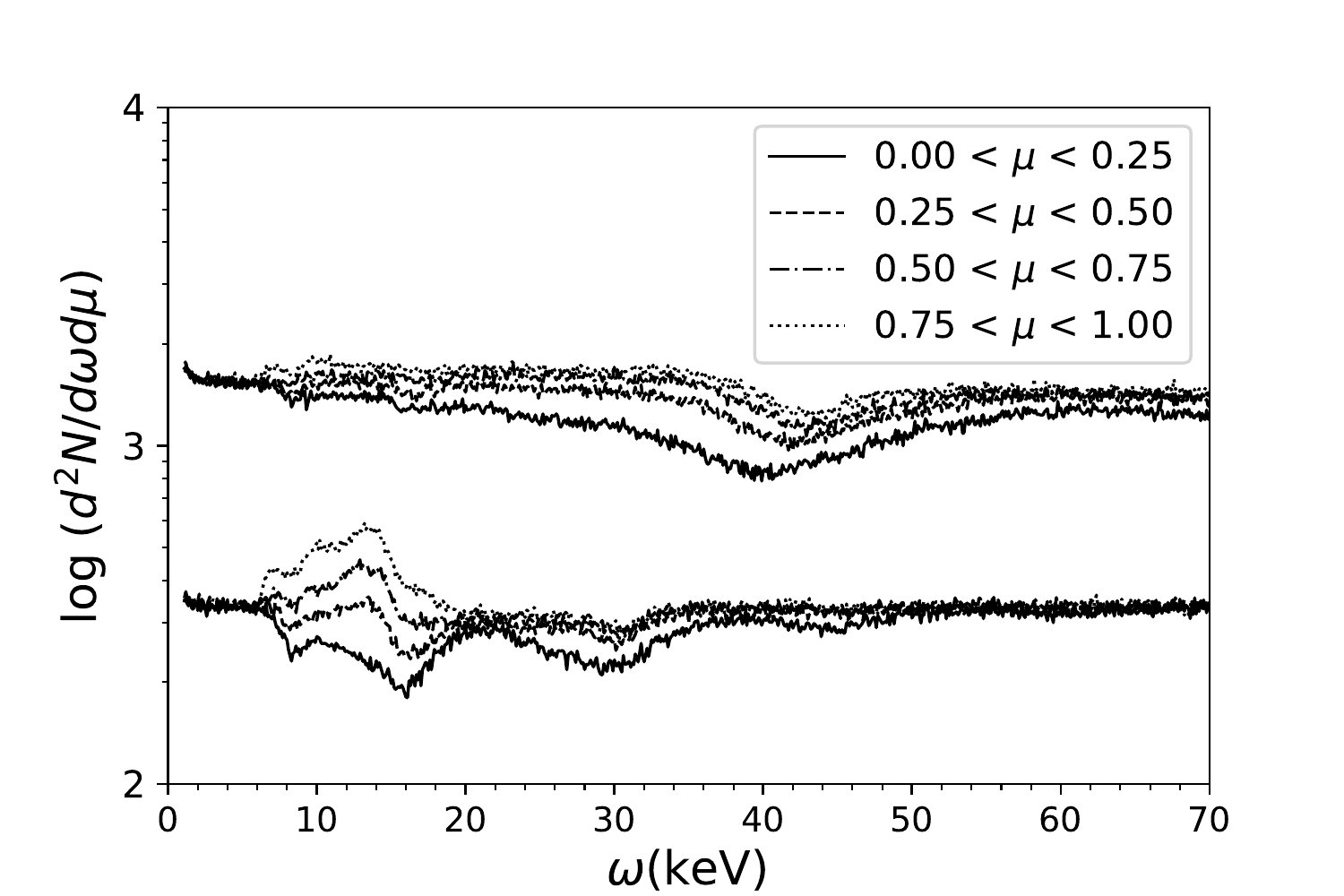}

\caption{\label{fig:f10}Spin phase averaged redshifted spectra without light
bending for flat continuum from two mounds of height 45m(bottom)
and 55m(top). The spectra are produced for a geometry of slab 1-0 with
uniform electron temperature $T_{e}=5\text{ KeV}$ and $\tau_{T}=10^{-3}$,
for four different $\mu$ bins. The density profile is derived from
the Paczynski equation of state \citep{Paczinsky_1983}.}
\end{figure}

\begin{figure}
\includegraphics[scale=0.55]{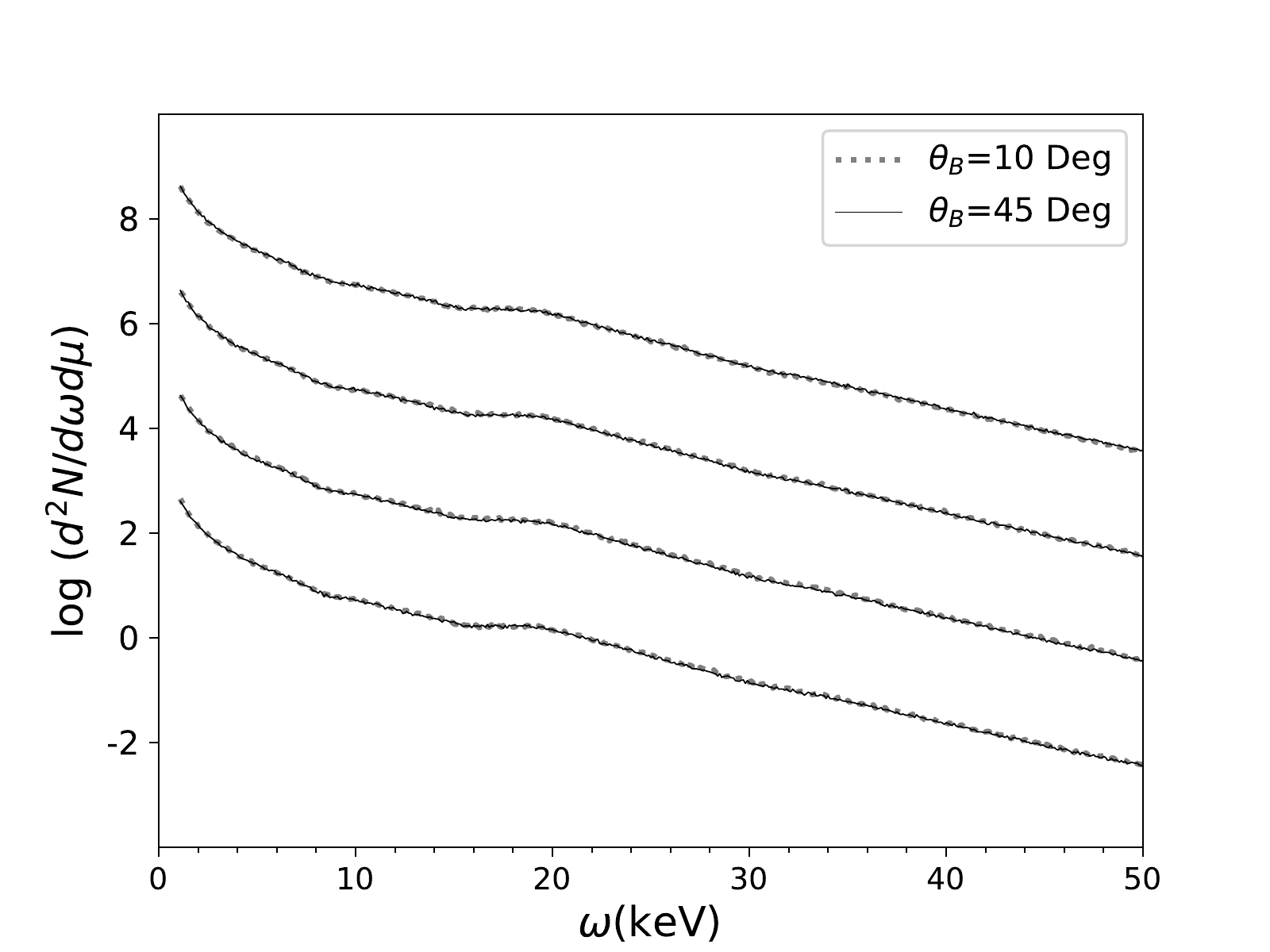}

\caption{\label{fig:11} {Spin phase averaged redshifted
spectra from 45m mound for \texttt{highecut} continuum including the light bending
for two angles $\theta_{B}=10^{\circ},45^{\circ}$ between the spin
axis and magnetic axis.  The spectra are plotted for four different
viewing bins $\mu_1$, $\mu_2$, $\mu_3$, $\mu_4$ from bottom to top. The input parameters are the same as in  Fig--\ref{fig:f10}.}}
\end{figure}

\begin{figure}
\includegraphics[scale=0.55]{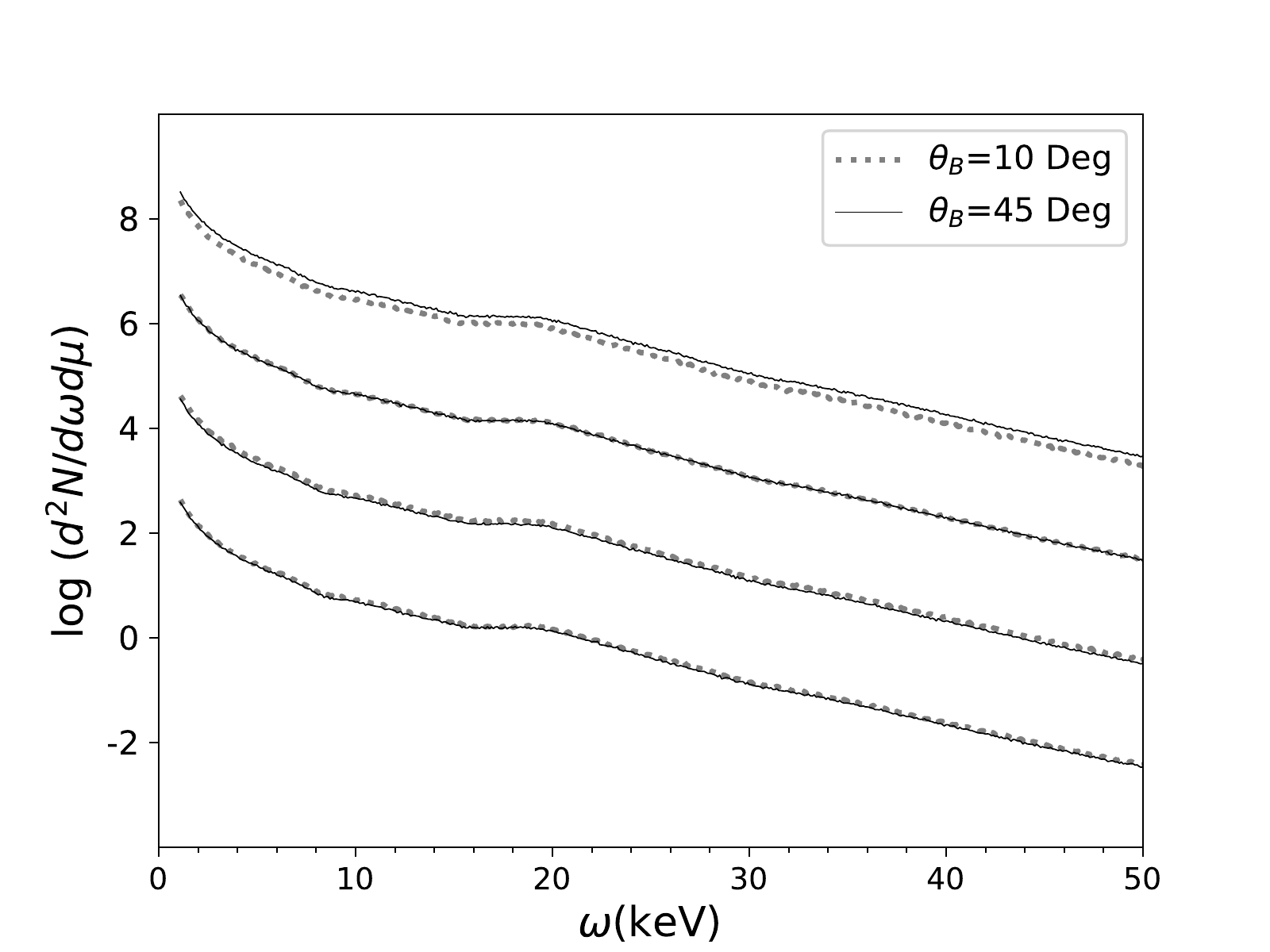}

\caption{\label{fig:m45_both_lb} {Same as Fig-\ref{fig:11} but including both the poles of the NS.} }
\end{figure}

The variation of $B_{r}$, $B_{z}$ and \textbf{ $B=\sqrt{B_{r}^{2}+B_{z}^{2}}$},
along the radial direction $r$ (in cylindrical co-ordinate system)
in the optically thin top layer of a mound, are shown in  Fig--\ref{fig:f9},
for two different heights, 45 m, and 55m. The variation in \textbf{$B_{r}$} is higher than that in $B_{z}$ along the radial direction $r$.
It can be seen that the maximum variation occurs at $r\sim700-800$ meter, and the maximum total field can reach
a value of $2-6$ times of the minimum.

The different panels (bottom and top) in  Fig--\ref{fig:f10} show
the redshifted spectra without light bending for mounds of height 45m and
55m, for different angle bins measured from the magnetic axis.
We find that at lower angles, the cyclotron lines have a small depth and at higher angular bins, the depth is much larger.
The width of the lines is found to be larger at all angles, compared to the width found for a constant magnetic field (Fig--\ref{fig:f5}).
The redshifted spectra from 45m mound (bottom of  Fig--\ref{fig:f10}), and at higher angular bins ($\mu_1$ and $\mu_2$) have cyclotron absorption features nearly at 7.7, 15.4, 30.8, 46.2 KeV. 
The pattern is primarily composed of two
sets of line feature. One, ($\zeta \omega_{n}^{cyc}(B^{'}\approx0.02)$)
from the region of the lower magnetic field $B^{'}\approx0.02$, contributing
at 7.7, 15.4, 23.1, 30.8, 38.5, ... KeV, and the second line pattern
($\zeta \omega_{n}^{cyc}(B^{'}\approx0.04)$) from near the maximum of the
magnetic field $B^{'}\approx0.04$, contributing at 15.4, 30.8, 46.2..
KeV. 
These two patterns overlap and create prominent absorption features
at 7.7, 15.4, 30.8, 46.2 KeV. Features at 23.1, 38.5 KeV, contributed
only by the region where $B^{'}=0.02$ and they are not deep enough to be vissible in
the composite spectrum.

Spectra from mounds with 55m height ( Fig--\ref{fig:f10}) share the
same qualitative features as those discussed for the 45m mound. The
magnetic distortion is larger in this case, and the line energies
are spread correspondingly further. The effective widths of the line
features are also significantly larger, particularly for features
originating in the high field regions. For the 55m mound the entire
spectrum is dominated by a single wide feature near 40 KeV (redshifted)
corresponding to the fundamental originating near the field maximum
of B'$\approx0.12$.

Clearly, the CRSF line energies produced from the accretion mounds do not follow the classical harmonic ratio for line features, as they are composed of lines due to different field strengths in different
regions.
Examples of anharmonic spacing of cyclotron features exist
also in observed sources. For example the phase averaged spectra of
4U0115+63 contain cyclotron features at 16.4, 23.2, 31.9, 48 KeV \citep{2002ApJ...580..394C}. However such anharmonicity is not easy to detect in
all cases due to the uncertainties in line energy estimation arising
from our lack of knowledge of the true underlying continuum.

Line features in observed sources are typically quite broad, for example
the width of the fundamental is $\sim6.4$ KeV for Her X-1, $\sim9$
KeV for 4U0352+309, $\sim8$ KeV for GX 301-2,$\sim7$ KeV for Cen
X-3 etc \citep{2002ApJ...580..394C}. In  Fig--\ref{fig:f10} the widths of
the first three harmonics are found to be nearly 3, 6, 7 KeV. So, the widths of features
obtained in our mound spectra are comparable to the observed values.

In  Fig--\ref{fig:11}, we present the spin phase averaged redshifted
spectra incorporating light bending for two angles $\theta_{B}=10^{\circ}$ and $45^{\circ}$,
between the spin axis and magnetic axis, and for two angle bins $\mu_1$ and $\mu_4$,
with respect to the spin axis. The angular averaging involved in this
causes the spectra for different angle bins to look very similar.
Relatively shallow spectral features near 7, 15 and 30 KeV are recognizable
in these spectra, as also in the spectra including both poles of the NS (Fig-\ref{fig:m45_both_lb}).

\subsection{Accretion column}
In this section we present the spectra produced by the accretion column for high luminosity sources. 
We take the plasma density profile of \citet{Becker_and_wolf_2007}. We assume that the mass accretion-rate is $\sim 10^{-8}M_{\odot}$/year, which gives a density of about $2\times10^{25}\text{ cm}^{-3}$ near the base (computed $\sim1$ cm above
the base) and $\sim2.7\times10^{20}\text{ cm}^{-3}$ at the top
of the column. We simulate an optically thin layer with perpendicular
Thompson optical depth $\tau_{T}$ of $10^{-3}$. Such a layer has
a thickness of about $10^{-5}\text{ cm}$ at the base and $\sim5\text{ cm}$
at the top.
For simplicity, we model the inner boundary of the simulation
volume as a truncated cone ( Fig--\ref{fig:f1} ). The dimenstions of
inner conical surface using the density profile of \citet{Becker_and_wolf_2007}
is computed in appendix A. Given the non-linear dependence of density
on $z$, $\tau_{T}$ (Eq--\ref{eq:a4}) to this conical surface somewhat
exceeds $10^{-3}$ except at the two ends. The photons are injected
at surface of this cone. The simulation region which is bounded by
the surface of the column and surface of the cone can be assumed to
consist of 
\begin{itemize}
\item 1) A thin slab at the top ($Sl$): It is situated at the top of the circular surface of the truncated cone, having height ( $\sim $5 cm) corresponding to $\tau_{T}=10^{-3}$.
The magnetic field intensity inside $Sl$ is assumed to be uniform ($B^{'}\simeq0.022$).
\item 2)  A near-cylindrical shell $Cy$: The region between the accretion column and the side walls of the truncated cone (Fig \ref{fig:f1}), having height $z_{b}=z_{c}-z_{t}$. 
Due to the large height of $Cy$, the variation of the magnetic field is noticeable,  $B^{'}\simeq0.03$ at the bottom and $B^{'}\simeq0.022$ at the top. 
\end{itemize}

\begin{figure}
\includegraphics[scale=0.6]{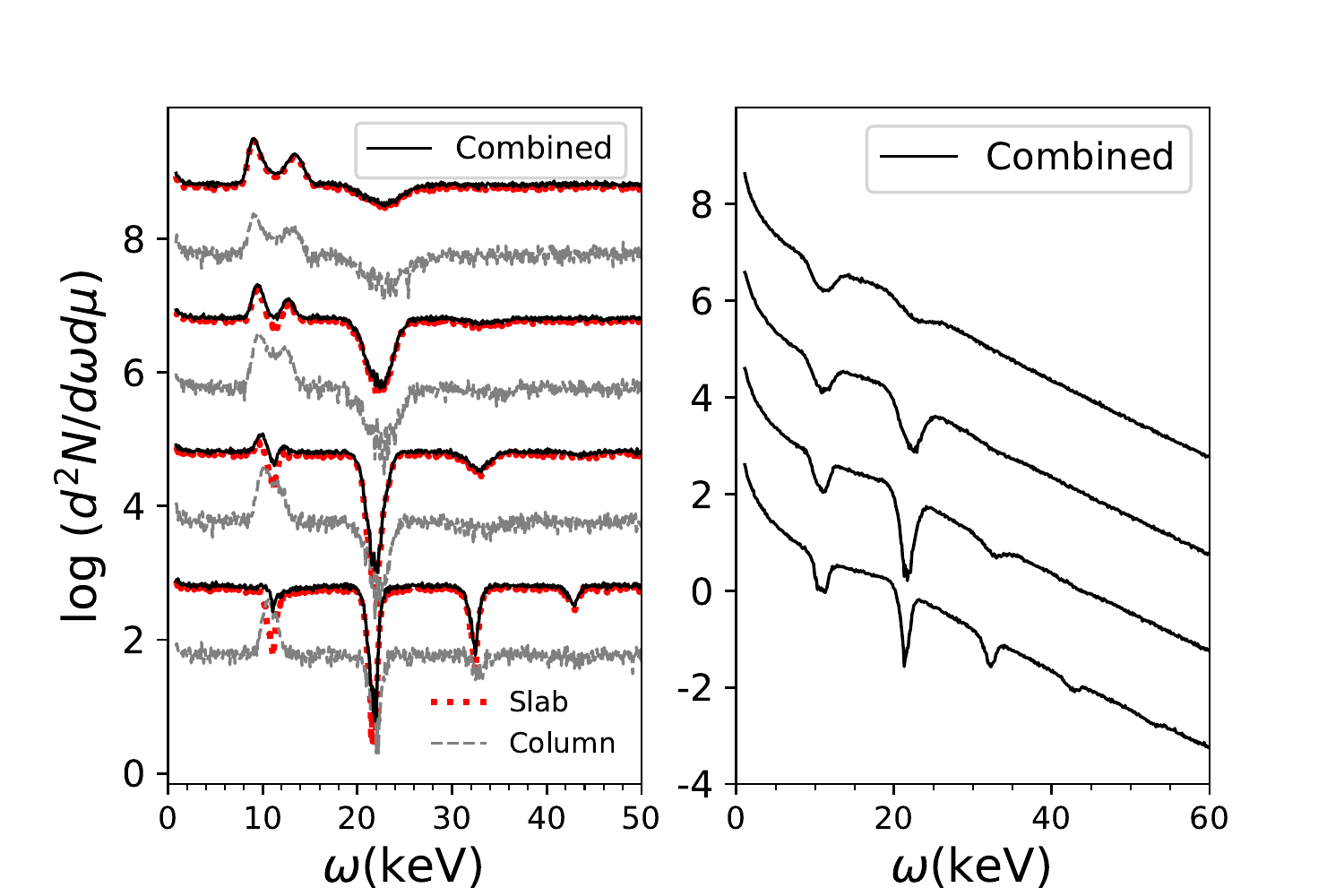}

\caption{ 
\label{fig:f15}Left panel: Spin phase averaged redshifted spectra
without light bending from accretion column for flat continuum in
four different angle bins (0.0$<\mu_1 \leq0.25$, $0.25<\mu_2\leq0.5$,  $0.5<\mu_3\leq0.75$, $0.75<\mu_4\leq1.0$). Red dotted lines: spectra from top slab ($Sl$),
Gray dashed lines: spectra from the sides ($Cy$), Black solid lines: total spectra
of $Sl$ and $Cy$. \protect \\
Right panel: Spin phase averaged redshifted spectra from accretion
column for \texttt{highecut} continnum in four different angle bins. The spectra
are the sum of those from $Sl$ and $Cy$ . 
}
\end{figure}

The left panel of the  Fig--\ref{fig:f15} shows the redshifted spectra for flat continuum. 
In this figure, the red dotted line corresponds to the output spectrum resulting from $Sl$, the gray dashed line corresponds to the output spectrum obtained from $Cy$, and the black solid curve represents the combined spectrum generated from $Sl$ and $Cy$. 
The spectra produced in $Sl$ are found to be qualitatively similar to those we found from a 1-0 slab with a constant field.
We find various harmonically spaced dips which are sharper at high angle bins and shallower at low angle bins. 
At low angle bins, a prominent presence of emission wings has also been found.
The variation in the field strength within $Cy$ broadens the line profile, and the line centres represent the average of the field intensity. 
In this case, we see two prominent features, an emission wing at 10 keV, and an absorption feature at 17 KeV.

\begin{figure}
\includegraphics[scale=0.55]{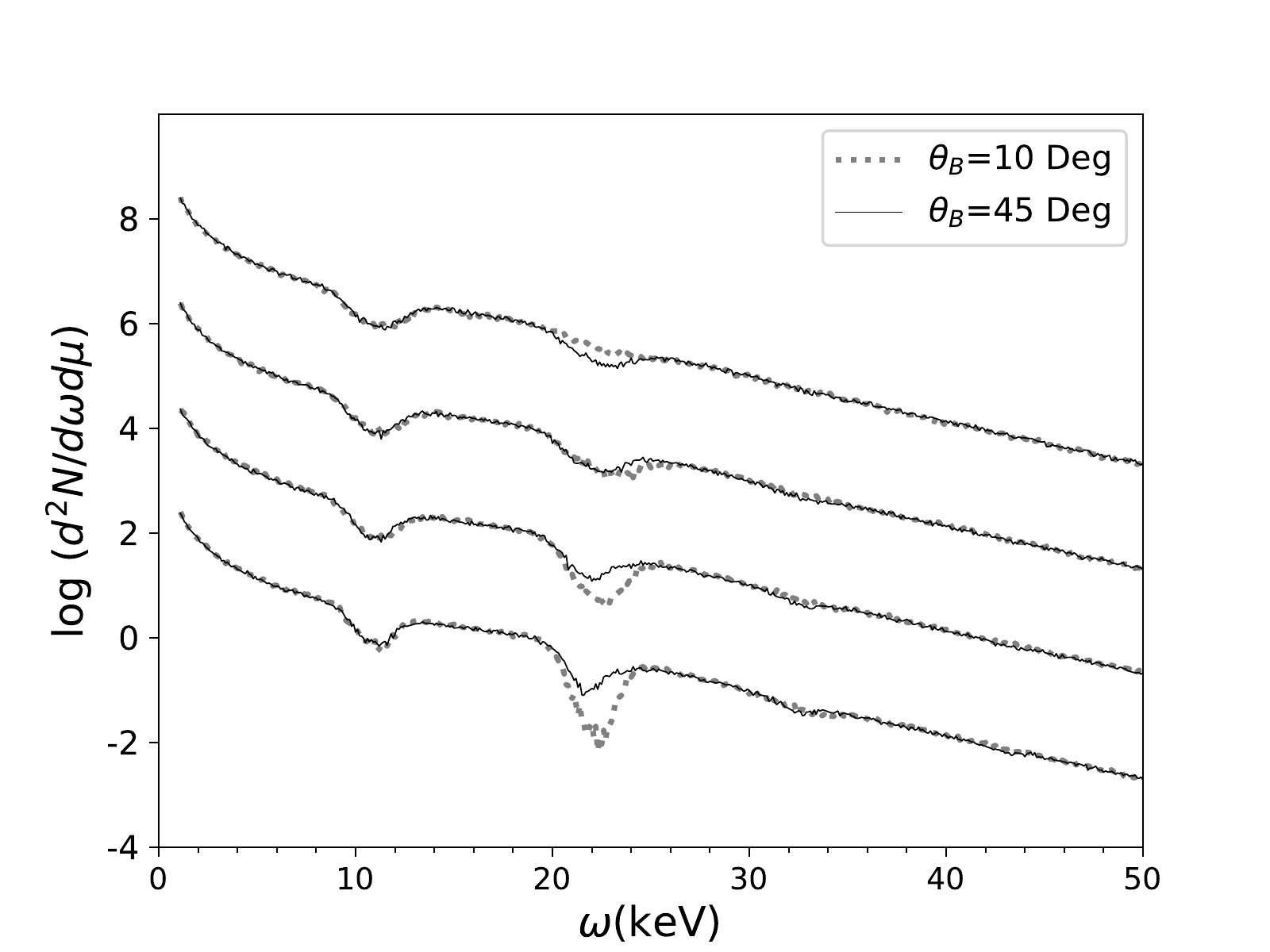}

\caption{ 
\label{fig:fig16}Spin phase averaged redshifted spectra from accretion
column including light bending for HCUT continnum for two angles
$\theta_{B}=0^{\circ},45^{\circ}$ between the spin axis and magnetic
axis. The spectra are the sum of those from $Sl$ and $Cy$, for four different angle bins (0.0$<\mu_1 \leq0.25$, $0.25<\mu_2\leq0.5$,  $0.5<\mu_3\leq0.75$, $0.75<\mu_4\leq1.0$).
}
\end{figure}



\begin{figure}
\includegraphics[scale=0.55]{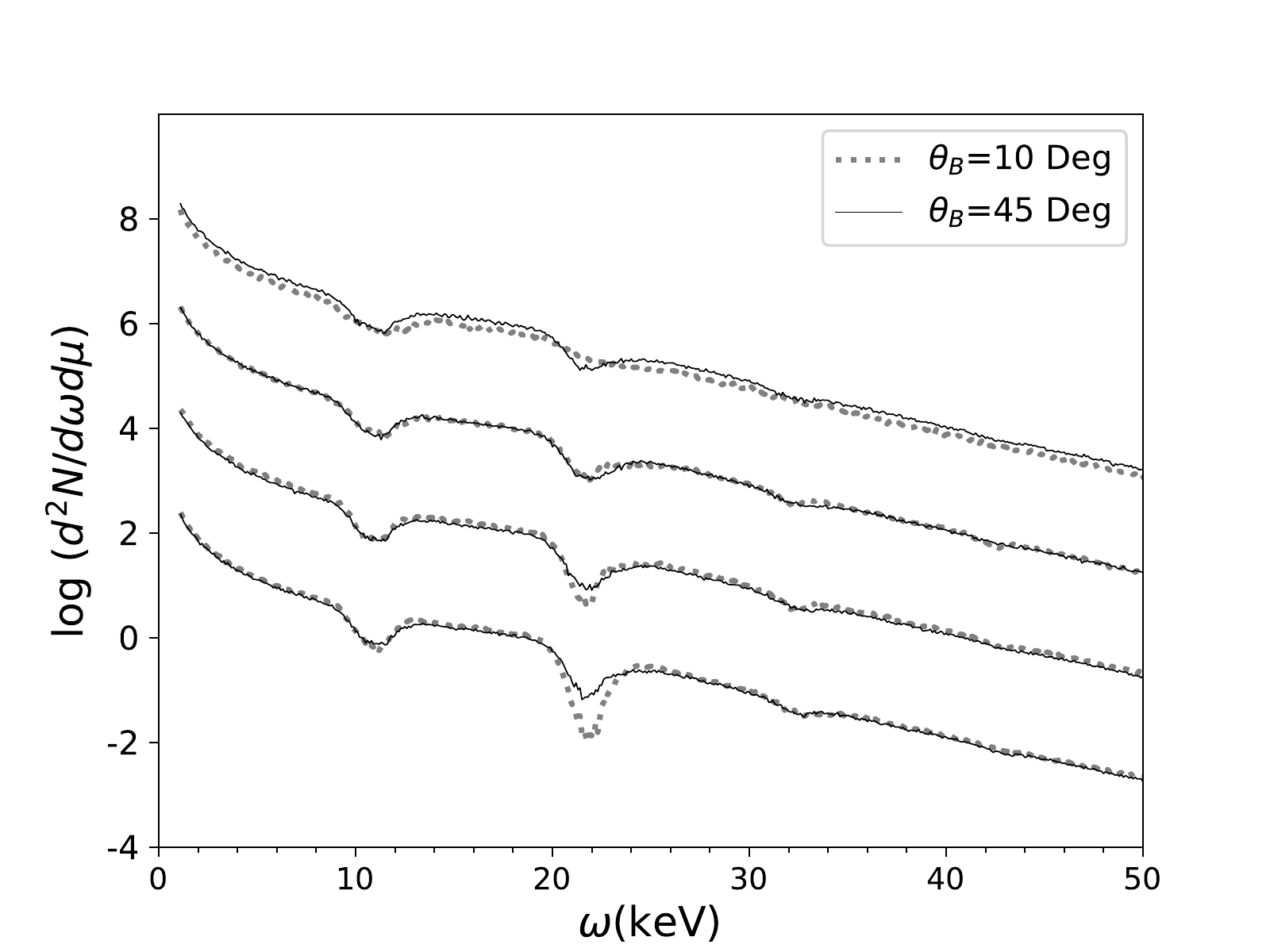}

\caption{ 
\label{fig:fig17} Same as Fig-\ref{fig:fig16}, including the contribution from both poles of the NS.
}
\end{figure}

The combined spectrum, obtained by summing the $Sl$ and the $Cy$ spectra,
have line features shallower than those for $Sl$, and the fundamental
appears in emission at all angle bins. Spectra generated
with \texttt{highecut} continuum are shown in the right panel of the  Fig--\ref{fig:f15}.
Due to the reduction in the number of transition photons, in this case the
emission features are suppressed. The spectra are found to depend significantly
on the angle from the magnetic axis. In the lowest angle bin $\mu_4$
the fundamental is deeper than the second harmonic, while at other
angle bins it is the opposite. At the highest angle bin $\mu_1$
up to five absorption features can be discerned.

Fig--\ref{fig:fig16} shows the spin phase averaged redshifted spectra for two values of $\theta_{B}$, $10^{\circ}$ and $45^{\circ}$, including the light-bending effects. For $\theta_{B}=10^{\circ}$, we find a variation of cyclotron line features with $\mu$, as
the effect of light bending at the top of $Cy$ is significantly
smaller than at the bottom (near NS surface).
For $\theta_{B}=45^{\circ}$, we find the spectra to be quite similar in all angle bins due to a greater angular averaging of the spectra. We can see the same effect  in the phase averaged combined spectra from both poles of the NS (Fig-\ref{fig:fig17}).

\section{Conclusions \label{sec:sec5}}

The main conclusions of our computations are summarized below.
\begin{itemize}
\item In the literature it is customary to compute the locally emergent
spectra as a function of the angle from the magnetic axis. We find
that the spin phase averaged spectra including general relativistic
effects, representing what would be seen by a distant observer, differ
substantially from the above. It is therefore essential to compute
the latter when a comparison with observed spectra have to be made.

\item The extent of angular averaging involved in the formation of observed
spectra depends on the amount of light bending, which is very severe
near the stellar surface, and reduces with altitude. If the emission
region is located very close to the stellar surface then the observed
spectra are rendered nearly independent of viewing angle unless the
local emergent spectra are very strongly anisotropic, as in the case
of a uniform field slab. If the emission region extends farther away
from the star, then the dependence on viewing angle becomes more perceptible.

\item The profiles of the cyclotron features, in particular of the fundamental,
are strongly dependent on the shape of the input continuum. The harder
the continuum, the more is the population of transition photons which
tend to fill in the fundamental feature and create emission wings.
This is seen very clearly in the spectra produced by a flat continuum.
On the other hand, for a power-law continuum with exponential cutoff the effect of transition photons
is minimal and the emission wings are virtually absent.

\item For a thin slab with uniform field, the local emergent spectra are
strongly dependent on the angle from the magnetic axis. Deep and narrow
features are seen at angles normal to the magnetic field, and the features are found to be broader and shallower
close to the magnetic axis. For a distant observer,
the wider features are found to dominate the spectra and harmonics above the second
are washed out. The line widths ($\sim2\text{ KeV}$) are found to be
relatively small compared to those usually seen in most X-ray pulsars. 

\item Spectra from accretion mounds show very broad and shallow lines with
overlapping features from subsequent harmonics. For a distant observer, up to three distinct features are visible. These features are
not necessarily harmonically spaced, and their widths reach up to
$\sim10\text{ KeV}$, not unlike some observed features but wider
than the average seen in most sources. 

\item The accretion column with dipole field produces spectral features
of intermediate width. They are asymmetric and somewhat anharmonically
spaced. The spectra viewed by a distant observer show detectable anisotropy,
but features higher than the second harmonic are washed out. 
\end{itemize}

The emission geometries we have explored in this work do generate features similar to those observed in some sources. But our simulations do not appear to reproduce the full range of observed behavior
in cyclotron spectra. 
In particular, more than three prominent harmonic
features have been observed in a few sources, like 4U0115+63, and in our simulation the higher harmonics are found to nearly disappear after applying the \texttt{highecut} continuum. This indicates a larger diversity
in the characteristics of the emission regions in different sources.
The existence of many prominent harmonic features and strong spin phase
dependence may suggest emission regions relatively far from the stellar
surface, and perhaps require more complex geometry than those explored
in this work. The Monte-Carlo code developed by us has the ability
to simulate many such complex situations, which will be attempted
in the future.

\section*{Acknowledgements}
The research work is funded by the the Council of Scientific and Industrial Research (CSIR), University Grants Commission (UGC) and Govt. of India. The authors want to thank the Inter University Centre for Astronomy and Astrophysics (IUCAA) for the research facilities. We sincerely acknowledge the use of high performance computing facilities at IUCAA, and in C-DAC, Pune. We want to thank the anonymous referee for the valuable comments and suggestions which  substantially improved the contents of the paper.

\section*{Data availability}
The data underlying this article will be shared on reasonable request to the corresponding author.

\bibliographystyle{mnras}
\bibliography{example} 

\begin{thebibliography}{}
\makeatletter
\relax
\def\mn@urlcharsother{\let\do\@makeother \do\$\do\&\do\#\do\^\do\_\do\%\do\~}
\def\mn@doi{\begingroup\mn@urlcharsother \@ifnextchar [ {\mn@doi@}
  {\mn@doi@[]}}
\def\mn@doi@[#1]#2{\def\@tempa{#1}\ifx\@tempa\@empty \href
  {http://dx.doi.org/#2} {doi:#2}\else \href {http://dx.doi.org/#2} {#1}\fi
  \endgroup}
\def\mn@eprint#1#2{\mn@eprint@#1:#2::\@nil}
\def\mn@eprint@arXiv#1{\href {http://arxiv.org/abs/#1} {{\tt arXiv:#1}}}
\def\mn@eprint@dblp#1{\href {http://dblp.uni-trier.de/rec/bibtex/#1.xml}
  {dblp:#1}}
\def\mn@eprint@#1:#2:#3:#4\@nil{\def\@tempa {#1}\def\@tempb {#2}\def\@tempc
  {#3}\ifx \@tempc \@empty \let \@tempc \@tempb \let \@tempb \@tempa \fi \ifx
  \@tempb \@empty \def\@tempb {arXiv}\fi \@ifundefined
  {mn@eprint@\@tempb}{\@tempb:\@tempc}{\expandafter \expandafter \csname
  mn@eprint@\@tempb\endcsname \expandafter{\@tempc}}}

\bibitem[\protect\citeauthoryear{{Alexander} \& {Meszaros}}{{Alexander} \&
  {Meszaros}}{1991}]{Alexander_meszaros_1991}
{Alexander} S.~G.,  {Meszaros} P.,  1991, \mn@doi [\apj] {10.1086/170001},
  \href {https://ui.adsabs.harvard.edu/abs/1991ApJ...372..565A} {372, 565}

\bibitem[\protect\citeauthoryear{{Alexander}, {Meszaros}  \&
  {Bussard}}{{Alexander} et~al.}{1989}]{Alexander_et_al_1989}
{Alexander} S.~G.,  {Meszaros} P.,   {Bussard} R.~W.,  1989, \mn@doi [\apj]
  {10.1086/167648}, \href
  {https://ui.adsabs.harvard.edu/abs/1989ApJ...342..928A} {342, 928}

\bibitem[\protect\citeauthoryear{{Araya} \& {Harding}}{{Araya} \&
  {Harding}}{1996}]{Araya_Harding_1996}
{Araya} R.~A.,  {Harding} A.~K.,  1996, \aaps, \href
  {https://ui.adsabs.harvard.edu/abs/1996A&AS..120C.183A} {120, 183}

\bibitem[\protect\citeauthoryear{{Araya} \& {Harding}}{{Araya} \&
  {Harding}}{1999}]{Araya_et_al_1999}
{Araya} R.~A.,  {Harding} A.~K.,  1999, \mn@doi [\apj] {10.1086/307157}, \href
  {https://ui.adsabs.harvard.edu/abs/1999ApJ...517..334A} {517, 334}

\bibitem[\protect\citeauthoryear{{Becker} \& {Wolff}}{{Becker} \&
  {Wolff}}{2007}]{Becker_and_wolf_2007}
{Becker} P.~A.,  {Wolff} M.~T.,  2007, \mn@doi [\apj] {10.1086/509108}, \href
  {https://ui.adsabs.harvard.edu/abs/2007ApJ...654..435B} {654, 435}

\bibitem[\protect\citeauthoryear{{Becker} et~al.,}{{Becker}
  et~al.}{2012}]{Becker_et_al_2012}
{Becker} P.~A.,  et~al., 2012, \mn@doi [\aap] {10.1051/0004-6361/201219065},
  \href {https://ui.adsabs.harvard.edu/abs/2012A&A...544A.123B} {544, A123}

\bibitem[\protect\citeauthoryear{{Beloborodov}}{{Beloborodov}}{2002}]{beloborodov02}
{Beloborodov} A.~M.,  2002, \mn@doi [\apjl] {10.1086/339511}, \href
  {http://adsabs.harvard.edu/abs/2002ApJ...566L..85B} {566, L85}

\bibitem[\protect\citeauthoryear{{Bonazzola}, {Heyvaerts}  \&
  {Puget}}{{Bonazzola} et~al.}{1979}]{Bonazola_et_al_1979}
{Bonazzola} S.,  {Heyvaerts} J.,   {Puget} J.~L.,  1979, \aap, \href
  {https://ui.adsabs.harvard.edu/abs/1979A&A....78...53B} {78, 53}

\bibitem[\protect\citeauthoryear{{Bulik}, {Meszaros}, {Woo}, {Hagase}  \&
  {Makishima}}{{Bulik} et~al.}{1992}]{Bulik_et_al_1992}
{Bulik} T.,  {Meszaros} P.,  {Woo} J.~W.,  {Hagase} F.,   {Makishima} K.,
  1992, \mn@doi [\apj] {10.1086/171676}, \href
  {https://ui.adsabs.harvard.edu/abs/1992ApJ...395..564B} {395, 564}

\bibitem[\protect\citeauthoryear{{Bulik}, {Riffert}, {Meszaros}, {Makishima},
  {Mihara}  \& {Thomas}}{{Bulik} et~al.}{1995}]{Bulik_et_al_1995}
{Bulik} T.,  {Riffert} H.,  {Meszaros} P.,  {Makishima} K.,  {Mihara} T.,
  {Thomas} B.,  1995, \mn@doi [\apj] {10.1086/175614}, \href
  {https://ui.adsabs.harvard.edu/abs/1995ApJ...444..405B} {444, 405}

\bibitem[\protect\citeauthoryear{{Coburn}, {Heindl}, {Rothschild}, {Gruber},
  {Kreykenbohm}, {Wilms}, {Kretschmar}  \& {Staubert}}{{Coburn}
  et~al.}{2002}]{2002ApJ...580..394C}
{Coburn} W.,  {Heindl} W.~A.,  {Rothschild} R.~E.,  {Gruber} D.~E.,
  {Kreykenbohm} I.,  {Wilms} J.,  {Kretschmar} P.,   {Staubert} R.,  2002,
  \mn@doi [\apj] {10.1086/343033}, \href
  {https://ui.adsabs.harvard.edu/abs/2002ApJ...580..394C} {580, 394}

\bibitem[\protect\citeauthoryear{{Freeman}, {Lamb}, {Wang}, {Wasserman},
  {Loredo}, {Fenimore}, {Murakami}  \& {Yoshida}}{{Freeman}
  et~al.}{1999}]{Freeman_et_al_1999}
{Freeman} P.~E.,  {Lamb} D.~Q.,  {Wang} J.~C.~L.,  {Wasserman} I.,  {Loredo}
  T.~J.,  {Fenimore} E.~E.,  {Murakami} T.,   {Yoshida} A.,  1999, \mn@doi
  [\apj] {10.1086/307818}, \href
  {https://ui.adsabs.harvard.edu/abs/1999ApJ...524..772F} {524, 772}

\bibitem[\protect\citeauthoryear{{Gil}, {Melikidze}  \& {Mitra}}{{Gil}
  et~al.}{2002}]{Gil_et_al_2002}
{Gil} J.~A.,  {Melikidze} G.~I.,   {Mitra} D.,  2002, \mn@doi [\aap]
  {10.1051/0004-6361:20020473}, \href
  {https://ui.adsabs.harvard.edu/abs/2002A&A...388..235G} {388, 235}

\bibitem[\protect\citeauthoryear{{Harding} \& {Daugherty}}{{Harding} \&
  {Daugherty}}{1991}]{Harding_Daugherty_1991}
{Harding} A.~K.,  {Daugherty} J.~K.,  1991, \mn@doi [\apj] {10.1086/170153},
  \href {https://ui.adsabs.harvard.edu/abs/1991ApJ...374..687H} {374, 687}

\bibitem[\protect\citeauthoryear{{Harding} \& {Preece}}{{Harding} \&
  {Preece}}{1987}]{Harding_Preece_1987}
{Harding} A.~K.,  {Preece} R.,  1987, \mn@doi [\apj] {10.1086/165510}, \href
  {https://ui.adsabs.harvard.edu/abs/1987ApJ...319..939H} {319, 939}

\bibitem[\protect\citeauthoryear{{Harding}, {Meszaros}, {Kirk}  \&
  {Galloway}}{{Harding} et~al.}{1984}]{Harding_et_al_1984}
{Harding} A.~K.,  {Meszaros} P.,  {Kirk} J.~G.,   {Galloway} D.~J.,  1984,
  \mn@doi [\apj] {10.1086/161801}, \href
  {https://ui.adsabs.harvard.edu/abs/1984ApJ...278..369H} {278, 369}

\bibitem[\protect\citeauthoryear{{Heindl}, {Rothschild}, {Coburn}, {Staubert},
  {Wilms}, {Kreykenbohm}  \& {Kretschmar}}{{Heindl}
  et~al.}{2004}]{Heindl_et_al_2004}
{Heindl} W.~A.,  {Rothschild} R.~E.,  {Coburn} W.,  {Staubert} R.,  {Wilms} J.,
   {Kreykenbohm} I.,   {Kretschmar} P.,  2004, in {Kaaret} P.,  {Lamb} F.~K.,
  {Swank} J.~H.,  eds,  American Institute of Physics Conference Series Vol.
  714, X-ray Timing 2003: Rossi and Beyond. pp 323--330 (\mn@eprint {arXiv}
  {astro-ph/0403197}), \mn@doi{10.1063/1.1781049}

\bibitem[\protect\citeauthoryear{{Isenberg}, {Lamb}  \& {Wang}}{{Isenberg}
  et~al.}{1998a}]{Isenberg_1998_b}
{Isenberg} M.,  {Lamb} D.~Q.,   {Wang} J. C.~L.,  1998a, \mn@doi [\apj]
  {10.1086/305124}, \href
  {https://ui.adsabs.harvard.edu/abs/1998ApJ...493..154I} {493, 154}

\bibitem[\protect\citeauthoryear{{Isenberg}, {Lamb}  \& {Wang}}{{Isenberg}
  et~al.}{1998b}]{Isenberg_1998_a}
{Isenberg} M.,  {Lamb} D.~Q.,   {Wang} J. C.~L.,  1998b, \mn@doi [\apj]
  {10.1086/306171}, \href
  {https://ui.adsabs.harvard.edu/abs/1998ApJ...505..688I} {505, 688}

\bibitem[\protect\citeauthoryear{{Latal}}{{Latal}}{1986}]{Latal_1986}
{Latal} H.~G.,  1986, \mn@doi [\apj] {10.1086/164609}, \href
  {https://ui.adsabs.harvard.edu/abs/1986ApJ...309..372L} {309, 372}

\bibitem[\protect\citeauthoryear{{Meszaros} \& {Nagel}}{{Meszaros} \&
  {Nagel}}{1985}]{Mesaros_Nagel_1985}
{Meszaros} P.,  {Nagel} W.,  1985, \mn@doi [\apj] {10.1086/163687}, \href
  {https://ui.adsabs.harvard.edu/abs/1985ApJ...299..138M} {299, 138}

\bibitem[\protect\citeauthoryear{{Meszaros}, {Nagel}  \& {Ventura}}{{Meszaros}
  et~al.}{1980}]{Meszaros_et_al_1980}
{Meszaros} P.,  {Nagel} W.,   {Ventura} J.,  1980, \mn@doi [\apj]
  {10.1086/158073}, \href
  {https://ui.adsabs.harvard.edu/abs/1980ApJ...238.1066M} {238, 1066}

\bibitem[\protect\citeauthoryear{{Meszaros}, {Harding}, {Kirk}  \&
  {Galloway}}{{Meszaros} et~al.}{1983}]{Meszaros_et_al_1983}
{Meszaros} P.,  {Harding} A.~K.,  {Kirk} J.~G.,   {Galloway} D.~J.,  1983,
  \mn@doi [\apjl] {10.1086/183973}, \href
  {https://ui.adsabs.harvard.edu/abs/1983ApJ...266L..33M} {266, L33}

\bibitem[\protect\citeauthoryear{{Mukherjee} \& {Bhattacharya}}{{Mukherjee} \&
  {Bhattacharya}}{2012}]{Mukherjee_and_Bhattacharya_2012}
{Mukherjee} D.,  {Bhattacharya} D.,  2012, \mn@doi [\mnras]
  {10.1111/j.1365-2966.2011.20085.x}, \href
  {https://ui.adsabs.harvard.edu/abs/2012MNRAS.420..720M} {420, 720}

\bibitem[\protect\citeauthoryear{{Nagel}}{{Nagel}}{1980}]{Nagel_1980}
{Nagel} W.,  1980, \mn@doi [\apj] {10.1086/157817}, \href
  {https://ui.adsabs.harvard.edu/abs/1980ApJ...236..904N} {236, 904}

\bibitem[\protect\citeauthoryear{{Nagel}}{{Nagel}}{1981a}]{Nagel_1981_b}
{Nagel} W.,  1981a, \mn@doi [\apj] {10.1086/159463}, \href
  {https://ui.adsabs.harvard.edu/abs/1981ApJ...251..278N} {251, 278}

\bibitem[\protect\citeauthoryear{{Nagel}}{{Nagel}}{1981b}]{Nagel_1981_a}
{Nagel} W.,  1981b, \mn@doi [\apj] {10.1086/159464}, \href
  {https://ui.adsabs.harvard.edu/abs/1981ApJ...251..288N} {251, 288}

\bibitem[\protect\citeauthoryear{{Nishimura}}{{Nishimura}}{2003}]{Nishimura_2003}
{Nishimura} O.,  2003, \mn@doi [\pasj] {10.1093/pasj/55.4.849}, \href
  {https://ui.adsabs.harvard.edu/abs/2003PASJ...55..849N} {55, 849}

\bibitem[\protect\citeauthoryear{{Nishimura}}{{Nishimura}}{2005}]{Nishimura_2005}
{Nishimura} O.,  2005, \mn@doi [\pasj] {10.1093/pasj/57.5.769}, \href
  {https://ui.adsabs.harvard.edu/abs/2005PASJ...57..769N} {57, 769}

\bibitem[\protect\citeauthoryear{{Nishimura}}{{Nishimura}}{2008}]{Nishimura_2008}
{Nishimura} O.,  2008, \mn@doi [\apj] {10.1086/523782}, \href
  {https://ui.adsabs.harvard.edu/abs/2008ApJ...672.1127N} {672, 1127}

\bibitem[\protect\citeauthoryear{{Nishimura}}{{Nishimura}}{2011}]{Nishimura_2011}
{Nishimura} O.,  2011, \mn@doi [\apj] {10.1088/0004-637X/730/2/106}, \href
  {https://ui.adsabs.harvard.edu/abs/2011ApJ...730..106N} {730, 106}

\bibitem[\protect\citeauthoryear{{Nishimura}}{{Nishimura}}{2019}]{Nishimura_2019}
{Nishimura} O.,  2019, \mn@doi [\pasj] {10.1093/pasj/psz008}, \href
  {https://ui.adsabs.harvard.edu/abs/2019PASJ...71...42N} {71, 42}

\bibitem[\protect\citeauthoryear{{Paczynski}}{{Paczynski}}{1983}]{Paczinsky_1983}
{Paczynski} B.,  1983, \mn@doi [\apj] {10.1086/160870}, \href
  {https://ui.adsabs.harvard.edu/abs/1983ApJ...267..315P} {267, 315}

\bibitem[\protect\citeauthoryear{{Payne} \& {Melatos}}{{Payne} \&
  {Melatos}}{2004}]{2004MNRAS.Payne&Melatos.351..569P}
{Payne} D.~J.~B.,  {Melatos} A.,  2004, \mn@doi [\mnras]
  {10.1111/j.1365-2966.2004.07798.x}, \href
  {http://adsabs.harvard.edu/abs/2004MNRAS.351..569P} {351, 569}

\bibitem[\protect\citeauthoryear{{Pottschmidt} et~al.,}{{Pottschmidt}
  et~al.}{2005}]{Pottschmidt_et_al_2005}
{Pottschmidt} K.,  et~al., 2005, \mn@doi [\apjl] {10.1086/498689}, \href
  {https://ui.adsabs.harvard.edu/abs/2005ApJ...634L..97P} {634, L97}

\bibitem[\protect\citeauthoryear{{Poutanen} \& {Gierli{\'n}ski}}{{Poutanen} \&
  {Gierli{\'n}ski}}{2003}]{poutanen03}
{Poutanen} J.,  {Gierli{\'n}ski} M.,  2003, \mn@doi [\mnras]
  {10.1046/j.1365-8711.2003.06773.x}, \href
  {http://adsabs.harvard.edu/abs/2003MNRAS.343.1301P} {343, 1301}

\bibitem[\protect\citeauthoryear{{Pravdo} \& {Bussard}}{{Pravdo} \&
  {Bussard}}{1981}]{Pravdo_Bussard_1981}
{Pravdo} S.~H.,  {Bussard} R.~W.,  1981, \mn@doi [\apjl] {10.1086/183566},
  \href {https://ui.adsabs.harvard.edu/abs/1981ApJ...246L.115P} {246, L115}

\bibitem[\protect\citeauthoryear{{Sch{\"o}nherr}, {Wilms}, {Kretschmar},
  {Kreykenbohm}, {Santangelo}, {Rothschild}, {Coburn}  \&
  {Staubert}}{{Sch{\"o}nherr} et~al.}{2007}]{Schonherr_et_al_2007}
{Sch{\"o}nherr} G.,  {Wilms} J.,  {Kretschmar} P.,  {Kreykenbohm} I.,
  {Santangelo} A.,  {Rothschild} R.~E.,  {Coburn} W.,   {Staubert} R.,  2007,
  \mn@doi [\aap] {10.1051/0004-6361:20077218}, \href
  {https://ui.adsabs.harvard.edu/abs/2007A&A...472..353S} {472, 353}

\bibitem[\protect\citeauthoryear{{Schwarm} et~al.,}{{Schwarm}
  et~al.}{2017a}]{2017A&A...597A...3S}
{Schwarm} F.~W.,  et~al., 2017a, \mn@doi [\aap] {10.1051/0004-6361/201629352},
  \href {https://ui.adsabs.harvard.edu/abs/2017A&A...597A...3S} {597, A3}

\bibitem[\protect\citeauthoryear{{Schwarm} et~al.,}{{Schwarm}
  et~al.}{2017b}]{2017A&A...601A..99S}
{Schwarm} F.~W.,  et~al., 2017b, \mn@doi [\aap] {10.1051/0004-6361/201630250},
  \href {https://ui.adsabs.harvard.edu/abs/2017A&A...601A..99S} {601, A99}

\bibitem[\protect\citeauthoryear{{Shabad}}{{Shabad}}{1975}]{Shabad_1975}
{Shabad} A.~E.,  1975, \mn@doi [Annals of Physics]
  {10.1016/0003-4916(75)90144-X}, \href
  {https://ui.adsabs.harvard.edu/abs/1975AnPhy..90..166S} {90, 166}

\bibitem[\protect\citeauthoryear{{Sina}}{{Sina}}{1996}]{Sina_1996}
{Sina} R.,  1996, PhD thesis, University of Maryland

\bibitem[\protect\citeauthoryear{{Sokolov} \& {Ternov}}{{Sokolov} \&
  {Ternov}}{1968}]{Sokolov_Ternov_1968}
{Sokolov} A.~A.,  {Ternov} I.~M.,  1968, Synchrotron radiation.
Akademie-Verlag, Berlin

\bibitem[\protect\citeauthoryear{{Staubert} et~al.,}{{Staubert}
  et~al.}{2019}]{2019A&A...622A..61S}
{Staubert} R.,  et~al., 2019, \mn@doi [\aap] {10.1051/0004-6361/201834479},
  \href {https://ui.adsabs.harvard.edu/abs/2019A&A...622A..61S} {622, A61}

\bibitem[\protect\citeauthoryear{{Truemper}, {Sacco}, {Pietsch}, {Reppin},
  {Kendziorra}  \& {Staubert}}{{Truemper} et~al.}{1977}]{1977MitAG..42..120T}
{Truemper} J.,  {Sacco} B.,  {Pietsch} W.,  {Reppin} C.,  {Kendziorra} E.,
  {Staubert} R.,  1977, Mitteilungen der Astronomischen Gesellschaft Hamburg,
  \href {http://adsabs.harvard.edu/abs/1977MitAG..42..120T} {42, 120}

\bibitem[\protect\citeauthoryear{{Wang}, {Wasserman}  \& {Salpeter}}{{Wang}
  et~al.}{1988}]{Wang_et_al_1988}
{Wang} J. C.~L.,  {Wasserman} I.~M.,   {Salpeter} E.~E.,  1988, \mn@doi [\apjs]
  {10.1086/191303}, \href
  {https://ui.adsabs.harvard.edu/abs/1988ApJS...68..735W} {68, 735}

\bibitem[\protect\citeauthoryear{{Yahel}}{{Yahel}}{1979}]{Yahel_1979}
{Yahel} R.~Z.,  1979, \mn@doi [\apjl] {10.1086/182933}, \href
  {https://ui.adsabs.harvard.edu/abs/1979ApJ...229L..73Y} {229, L73}

\makeatother
\end{thebibliography}




\bsp	

\appendix

\section{Estimates of dimension of the Conical source surface}
Here we have estimated the dimensions of the conical injection surface and
the volume in which radiative transfer take place in an accretion column (Case-III, Fig--\ref{fig:f1}).
We get a Thompson optical depth $\tau_{T}(l)$ by integrating on density $n_{e}(z)$ (Eq--\ref{eq:4.1.6}) over depth
$l$. Where, 
$l$ is measured from the top of the column.
In terms of altitude $z$ from the base, $l=z_c-z$, where $z_c$ is the total height of the column (Fig \ref{fig:f1}).
We compute the Thompson optical depth as a function
of altitude $\tau_{T}(z)$. An inversion is then performed
to get the height $z_{T}$ as a function of Thompson optical depth
$\tau_{T}$ . 

\begin{equation}
z_{T}(\tau_{T})=\dfrac{z_{c}}{ln(a)}ln\left(1+(a-1)\exp\left(-\dfrac{\tau_{T}\ln(a)}{n_{0}\sigma_{T}z_{c}}\right)\right)\label{eq:a4}
\end{equation}
where $a=7/3$, $z$ is measured from the bottom of the column and
$n_{0}$ is a parameter having the dimension of number density:

\begin{equation}
n_{0}=\dfrac{\dot{M_{c}}}{\pi r_{c}^{2}v_{ff}m_{He}}\label{eq:a3}
\end{equation}
The height of the truncated cone can be computed for Thompson optical
depth $\tau_{T}=10^{-3}$ using Eq--\ref{eq:a4}

\begin{equation}
z_{b}=z_{T}(\tau_{T}=10^{-3})\label{eq:a5}
\end{equation}
 Next, we have estimated the radius of the upper and lower circular
base ($r_{t}$,$r_{b}$ , with horizontal Thompson optical depth $\tau_{t}=10^{-3}$) of the cone (Fig--\ref{fig:f1}). While computing ($r_{t}$,$r_{b}$) we assume that the density $n_{e}$ does not vary along the horizontal
direction.
The thickness of the optically thin
layer $r_{tt}$ at the height $z_{b}$ and,
$r_{bb}$ at the bottom of the column can be computed as

\begin{equation}
r_{tt}=\dfrac{\tau_{T}(=10^{-3})}{n_{e}(z_{b})\sigma_{T}},\,\,\,\,\, r_{bb}=\dfrac{\tau_{T}(=10^{-3})}{n_{e}(z_{\epsilon})\sigma_{T}}\label{eq:a6}
\end{equation}
 where $z_{\epsilon}$ is a very small height above the base. Note
that the $n_{e}(z)\rightarrow\infty$ as $z\rightarrow0$ so to we
have taken $z_{\epsilon}=1\text{ cm}$, where the value of $r_{bb}$
is of order $10^{-3}\text{cm}$. The radius of the top and bottom
circular base of the cone can be computed as (Fig--\ref{fig:f1}),

\begin{equation}
r_{t}=r_{c}-r_{tt},\,\,\,\,\,\, r_{b}=r_{c}-r_{bb}\label{eq:a8}
\end{equation}
The half angle of the apex of cone is then

\begin{equation}
\theta_{h}=\tan^{-1}((r_{b}-r_{t})/z_{b}))\label{eq:a9}
\end{equation}
The next step is to inject photons isotropically on the surface of
the cone. The PDF of isotropic photon injection is given by the fractional
area. After using the inverse function method we can get equations
for $r$ and $\phi$ as,

\begin{equation}
r=\sqrt{\left(r_{t}^{2}+\left(r_{b}^{2}-r_{t}^{2}\right)\xi\right)},\,\,\,\,\,\,\phi=2\pi\xi\label{eq:a12}
\end{equation}
where $\xi$ is an uniform variate. The altitude $z$ corresponding
to $r$ can be obtained from the geometry of the cone, 

\begin{equation}
z=(r_{b}-r)/\text{tan}(\theta_{h})\label{eq:a14}
\end{equation}
so the photon injection is performed at the position 

\begin{equation}
x_{inj}=r\cos(\phi),\,\,\, y_{inj}=r\sin(\phi),\,\,\, z_{inj}=z\label{eq:a15}
\end{equation}
We select the propagation angle $\theta_{inj},\phi_{inj}$
for the photon. For the injection of a photon at $x_{inj},y_{inj},z_{inj}$
in a certain direction $\theta_{inj},\phi_{inj}$ we consider a tangent
plane at the point $x_{inj},y_{inj},z_{inj}$. The photons are injected outwards of the tangent plane, and away from the surface of the cone. The injection angles are chosen first in a local co-ordinate
system and then transformed to the global co-ordinate system. The
global co-ordinate system has its origin at the center of the base of the column,
and its $z$ axis is aligned with the magnetic axis. The local co-ordinate
system has its origin at $x_{inj},y_{inj},z_{inj}$ and its $z$ axis
is along the slant surface, pointing towards the apex. The $x$ axis of the local co-ordinate system
is chosen to be perpendicular to the tangent plane, and pointed towards the outward direction.
Two rotations by angles $\theta_{h},\phi$ are needed to transform
the propagation angles from the local frame to the global frame. In
the local frame the angle $\theta_{inj}^{''}$ is chosen in the
range $(0,\pi)$, but $\phi_{inj}^{''}$ is chosen to be within the range
($-\pi/2$,$\pi/2$) to ensure the outward propagation of the photon. The direction cosines corresponding to $\theta_{inj}^{''}$,
$\phi_{inj}^{''}$ in the local frame are,

\begin{equation}
\begin{array}{c}
\Omega_{x}^{''}=\sin\theta_{inj}^{''}\cos\phi_{inj}^{''}\\
\Omega_{y}^{''}=\sin\theta_{inj}^{''}\sin\phi_{inj}^{''}\\
\Omega_{x}^{''}=\cos\theta_{inj}^{''}
\end{array}\label{eq:a16}
\end{equation}
which may be is transformed to the global frame via two rotations,

\begin{equation}
\left(\begin{array}{c}
\Omega_{x}\\
\Omega_{y}\\
\Omega_{z}
\end{array}\right)=\left(\begin{array}{ccc}
\cos(\theta_{h})\cos(\phi) & -\sin(\phi) & -\sin(\theta_{h})\cos(\phi)\\
\cos(\theta_{h})\sin(\phi) & \cos(\phi) & -\sin(\theta_{h})\sin(\phi)\\
\sin(\theta_{h}) & 0 & \cos(\theta_{h})
\end{array}\right)\left(\begin{array}{c}
\Omega_{x}^{''}\\
\Omega_{y}^{''}\\
\Omega_{z}^{''}
\end{array}\right)\label{eq:a17}
\end{equation}
from which we compute the injection angles $\theta_{inj},\phi_{inj}$
:

\[
\theta_{inj}=\tan^{-1}\left(\dfrac{\sqrt{\Omega_{x}^{2}+\Omega_{y}^{2}}}{\Omega_{z}}\right)
\]

\[
\phi_{inj}=\tan^{-1}\left(\dfrac{\Omega_{y}}{\Omega_{x}}\right)
\]

\begin{figure}
\includegraphics[scale=0.33]{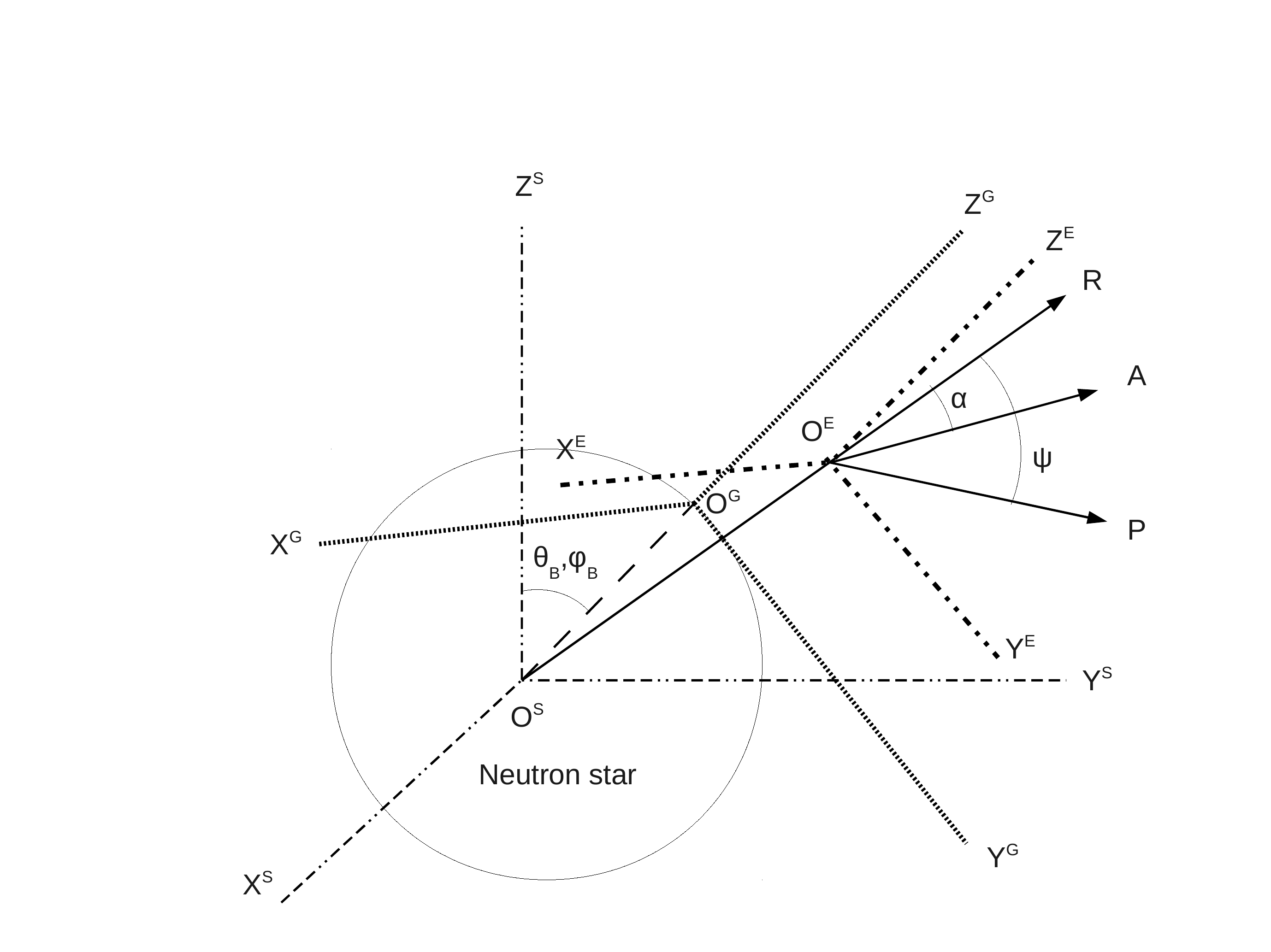}

\caption{\label{fig:appene1}The figure shows schematic diagram used to compute
the light bending. The $St$ frame has the center at $O^{s}$, the
$Gb$ frame has the center at $O^{G}$ and the $Es$ frame has the
center at $O^{E}$.}
\end{figure}

\section{Light bending}

The photon trajectory near the compact object bends due the strong
gravitational field. The photon which is emitted
at the point $R_{x}$,$R_{y}$,$R_{z}$ ($R\boldsymbol{\hat{R}})$
(measured from center of compact object) at an angle $\alpha$ measured
from the radial direction $\boldsymbol{\hat{R}}$, reaches the observer
at infinity at an angle $\psi$ (measured from the radial direction
$\boldsymbol{\hat{R}}$). A very simple but approximate formula of
gravitational light bending has been derived by \citet{beloborodov02}.

\begin{equation}
1-\text{cos}\alpha=(1-\text{cos}\psi)\left(1-\dfrac{r_{g}}{R}\right)\label{eq:appene1}
\end{equation}
where $r_{g}=2GM/c^{2}$ is the Schwarzchild radius of the neutron
star. The formula is valid up to $\alpha=90^{\circ}$, and beyond that
the error grows rapidly. If the unit vectors along the direction of
emission ($\alpha$ measured from the radial direction) of a photon
at the neutron star surface and the direction of escape of the photon
to the observer ($\psi$ measured from the radial direction ) are
$\boldsymbol{n}_{\alpha}$,$\boldsymbol{n}_{\psi}$ respectively, then from \citet{poutanen03} we get,

\begin{equation}
\boldsymbol{n}_{\psi}=\dfrac{\sin\psi}{\sin\alpha}\boldsymbol{n}_{\alpha}-\dfrac{\sin(\psi-\alpha)}{\sin\alpha}\boldsymbol{\hat{R}}\label{eq:appene2}
\end{equation}
 Fig--\ref{fig:appene1} shows the schematic diagram of light bending and the geometry.
 \begin{itemize}
     \item i) $St$($X^{s},Y^{s},Z^{s}$): The reference frame
with origin $O^{S}$ at the center of the star with its z axis along
the spin axis of the neutron star.

\item ii) $Gb$($X^{G},Y^{G},Z^{G}$): The reference frame 
with its origin at the center of the bottom circular surface $O^{G}$
of the accretion column with its z axis parallel to the magnetic axes
of the neutron star.

\item iii)  $Es$($X^{E},Y^{E},Z^{E}$): The reference frame
with origin $O^{E}$ at the point where photon escapes and all its
axes parallel to those of the frame $Gb$. 
 \end{itemize}

We compute the light
bending in a local frame $Es$ (light bending angle is the same in
the $Gb$ frame since the axes of $Es$ and $Gb$ are parallel) and
then the direction of light bending has been obtained in the frame $St$.
To compute the light bending in frame $Es$ we have represented the
vectors in Eq--\ref{eq:appene2}, in the local frame $Es$. Let $\boldsymbol{n}_{\alpha}=\boldsymbol{O^{E}A}/|\boldsymbol{O^{E}A}|$
be the direction at which the photon emerges, $\boldsymbol{n}_{\psi}=\boldsymbol{O^{E}B}/|\boldsymbol{O^{E}A}|$
be the direction in which photon reaches the observer after light
bending, and $\hat{\boldsymbol{R}}=\boldsymbol{O^{s}R}/|\boldsymbol{O^{S}A}|$ be the radial vector from the center of the neutron star to the emergence point. The photons are injected in the frame $Gb$ at angle $\theta_{i}$,$\phi_{i}$
measured from the $Z^{G}$ axis so 

\begin{equation}
\begin{array}{c}
n_{\alpha x}=\sin\theta_{i}\cos\phi_{i}\\
n_{\alpha y}=\sin\theta_{i}\sin\phi_{i}\\
n_{\alpha z}=\cos\theta_{i}
\end{array}\label{eq:appene3}
\end{equation}
 $\boldsymbol{\hat{R}}$ can be written as 

\begin{equation}
\hat{\boldsymbol{R}}=\dfrac{\boldsymbol{R_{*}}+\boldsymbol{O^{G}O^{E}}}{|\boldsymbol{R_{*}}+\boldsymbol{O^{G}O^{E}}|}\label{eq:appene4}
\end{equation}
$\boldsymbol{\hat{R}}$ can be represented in the frame $Gb$. The
angle $\alpha$ can now be obtained with the dot product of $\boldsymbol{\hat{R}}$,\textbf{$\mathbf{n}_{\alpha}$}

\begin{equation}
\alpha=\cos^{-1}(\mathbf{\hat{R}}\bullet\boldsymbol{n}_{\alpha})\label{eq:appene5}
\end{equation}
 using the values of $\boldsymbol{n}_{\alpha}$, $\boldsymbol{\hat{R}}$,
$\alpha$ and putting the value of $\psi$ from Eq--\ref{eq:appene1}
in terms of $\alpha$ in Eq--\ref{eq:appene2} one can get the value
of the components of the unit vector of light bending $\boldsymbol{n}_{\psi}$
in the $Gb$ frame. This vector which is computed in the frame $Gb$
can be transformed to the frame $St$ by making two rotations in angle
$\theta_{B}$,$\phi_{B}$. Here $\theta_{B}$,$\phi_{B}$ are the polar and azimuth angle of magnetic field axis ($O^GZ^G$) in $S_t$ frame.





\label{lastpage}
\end{document}